\documentclass[iop]{emulateapj}
\usepackage[english]{babel}     
\usepackage{enumitem}
\usepackage{amssymb,amsmath}    
\usepackage{graphicx}                   
\usepackage{braket}             
\usepackage{hyperref}
\usepackage{natbib}
\usepackage{bm}
\usepackage{multirow}

\shorttitle{CG~611} 
\shortauthors{Graham et al.}

\begin{document}

\title{Implications for the origin of dwarf early-type galaxies: a detailed
  look at the isolated rotating dwarf early-type galaxy LEDA 2108986 (CG 611),
ramifications for the Fundamental Plane's $S^2_K$ kinematic scaling and the spin--ellipticity diagram}

\author{Alister W.\ Graham, Joachim Janz}
\affil{Centre for Astrophysics and Supercomputing, Swinburne University of
  Technology, Victoria 3122, Australia.}
\email{AGraham@astro.swin.edu.au}

\author{Samantha J.\ Penny}
\affil{Institute of Cosmology and Gravitation, University of Portsmouth,
  Dennis Sciama Building, Burnaby Road, Portsmouth PO1 3FX, UK.}

\author{Igor V.\ Chilingarian}
\affil{Smithsonian Astrophysical Observatory, 60 Garden Street MS09, Cambridge, MA
02138, USA.; Sternberg Astronomical Institute, Moscow State University, 13
Universitetsky prospect, Moscow 119992, Russia.}

\author{Bogdan C.\ Ciambur, Duncan A.\ Forbes}
\affil{Centre for Astrophysics and Supercomputing, Swinburne University of
  Technology, Victoria 3122, Australia.}

\author{Roger L.\ Davies}
\affil{Sub-department of Astrophysics, Department of Physics, University of Oxford,
Denys Wilkinson Building, Keble Road, Oxford OX1 3RH, UK.}

\begin{abstract}

Selected from a sample of nine, isolated, dwarf early-type galaxies (ETGs)
with the same range of kinematic properties as dwarf ETGs in clusters, we
use LEDA~2108986 (CG~611) to address the Nature versus Nurture debate
regarding the formation of dwarf ETGs.  The presence of faint disk structures
and rotation within some cluster dwarf ETGs has often been heralded as
evidence that they were once late-type spiral or dwarf irregular galaxies
prior to experiencing a cluster-induced transformation into an ETG.  However,
CG~611 also contains significant stellar rotation ($\approx$20~km~s$^{-1}$)
over its inner half-light radius ($R_{\rm e,maj}=0.71$~kpc), and its stellar
structure and kinematics resemble those of cluster ETGs.  In addition to
hosting a faint young nuclear spiral within a possible intermediate-scale
stellar disk, CG~611 has accreted an intermediate-scale, counter-rotating gas
disk.  It is therefore apparent that dwarf ETGs can be built by accretion
events, as opposed to disk-stripping scenarios.  We go on to discuss how both
dwarf and ordinary ETGs with intermediate-scale disks, whether under
(de)construction or not, are not fully
represented by the kinematic scaling $S_{0.5}=\sqrt{0.5\,V^2_{\rm
    rot}+\sigma^2}$, and we also introduce a modified spin--ellipticity diagram
$\lambda(R)$--$\epsilon(R)$ with the potential to track galaxies with such
disks.

\end{abstract}


\section{Introduction}

Dwarf early-type\footnote{Throughout this paper, the terms ``early-type'' and
  ``late-type'' are used in reference to a galaxy's stellar morphology, rather
  than its stellar population.} galaxies (ETGs) --- encompassing dwarf
elliptical (dE), dwarf ellicular\footnote{The term ``ellicular'' was
  introduced in Graham et al.\ (2016) to describe the ``ES'' galaxy class
  (intermediate between E and S0) that was introduced by Liller (1966).}
(dES), and dwarf lenticular (dS0) galaxies --- are commonly found in clusters
of galaxies but are relatively scarce in the field.  Building on van den Bergh
(1976) and Butcher \& Oemler (1978), it has been theorized that these low-mass
galaxies were originally late-type spiral galaxies or irregular disk galaxies
that were stripped of their gas (Gunn \& Gott 1972; Larson et al.\ 1980) and
then ``harassed'' and reshaped into ETGs by either a cluster environment
(Moore et al.\ 1996, 1998; Mastropietro et al.\ 2005) or through ``tidal
stirring'' by a massive neighbor (Mayer et al.\ 2001a, 2001b).  While there is
abundant observational evidence of gas-stripping (e.g.\ Gavazzi et al.\ 2001;
Michielsen et al.\ 2008; Yagi et al.\ 2010; Owers et al.\ 2012; Merluzzi et
al.\ 2013, 2016; Ebeling et al.\ 2014; Boselli et al.\ 2016), here we
investigate the case for morphological transformation, specifically, whether
rotation and (weak) disk structures in dwarf ETGs are actually proof of
late-type galaxies transformed into dwarf ETGs --- as is commonly advocated in
the literature (e.g.\ Boselli et al.\ 2008; De Rijcke et al.\ 2010; Kormendy
\& Bender 2012; Penny et al.\ 2014; Benson et al.\ 2015; Ry\'s et al.\ 2015; 
Toloba et al.\ 2015; Janz et al.\ 2016).
 
Dwarf ETGs have typically been distinguished by low luminosities ($M_B > -18$
mag; $M_r \gtrsim -19.3$ mag) and low stellar masses ($M_* < 5\times
10^9~M_{\odot}$), and the absence of wide-spread star formation.  Furthermore,
their lack of a large-scale spiral pattern voids their membership of the
spiral, i.e.\ the late-type, galaxy class.  A key reasoning in support of the
galactic metamorphosis of ETGs from late-type galaxies has been the presence
of a significant rotational component (e.g.\ Pedraz et al.\ 2002; Simien \&
Prugniel 2002; Geha et al.\ 2003; de Rijcke et al.\ 2005; Chilingarian et
al.\ 2009; Geha et al.\ 2010; Toloba et al.\ 2015; Penny et al.\ 2016) which
has typically been interpreted as a remnant of the initial disk from which
these galaxies were supposedly transformed.  However, most ordinary,
i.e.\ non-dwarf, ETGs are also now known to contain a substantial rotating
disk\footnote{ETGs brighter than $M_B \approx -21.25$ mag, or with stellar
  masses greater than $\approx 10^{11}~M_{\odot}$, tend not to have disks.}.
That is, they are not the purely pressure supported systems held up by random
stellar motions that they were once considered to be.   
Measuring the major-axis kinematics of the brightest ETGs in
the Fornax cluster, Graham et al.\ (1998) reported that ``the true number of
{\it dynamically hot} stellar systems is much lower than previously thought,'' 
a result also seen using 2D kinematic maps of Fornax ETGs (Scott et al.\ 2014)
and confirmed in notably larger samples of ETGs beyond the Fornax cluster
(Emsellem et al.\ 2011; Oh et al.\ 2016).  However, many of these galaxies are
too massive for ``harassment'' (of what was originally a spiral galaxy) to
have been effective, and many are not even located in a cluster environment
where ``harassment'' could have played its (harsh) nurturing hand.

If the rotating disks in ordinary ETGs have been built
by accretion events, from either an initial primordial collapse 
and/or around a pre-existing spheroidal component (Graham 
et al.\ 2015, 2016; de la Rosa et al.\ 2016), then it somewhat undermines the above
argument for a transformation scenario for the dwarf ETGs. 
Moreover, the kinematic properties of dwarf 
and ordinary ETGs display no sign of a divide at $M_B = -18$ mag (e.g.\ Norris
et al.\ 2014; Penny et al.\ 2016) --- which is the 
magnitude commonly used for the dwarf/ordinary naming convention but is also alleged to mark a
separation about which distinct physical processes operate 
(Kormendy \& Bender 2012, and references therein).  Rather than a dichotomy of
merger-built ordinary ETGs versus ``harassed'' late-type galaxies that have
been transformed into dwarf ETGs, there is a continuity at $M_B = -18$ mag of
physical properties such as the presence of disks, the occurrence of kinematic
substructure, metallicity, metallicity gradients, globular cluster systems,
mass-to-light ratios, etc.\ (e.g.\ Forbes et al.\ 1996, 2008; 
Graham \& Guzm\'an 2003; Gavazzi et
al.\ 2004; de Rijcke et al.\ 2005; Ferrarese et al.\ 2006; C\^ot\'e et
al.\ 2007, 2008; Janz \& Lisker 2008, 2009; den
Brok et al.\ 2011; Koleva et al.\ 2011; Weinmann et al.\ 2011; Ferrarese 2016).  The 
luminosity-velocity dispersion relation is also continuous at $M_B = -18$ mag,
with a logarithmic slope of 2 (e.g.\ Davies et al.\ 1983; Held et al.\ 1992;
de Rijcke et al.\ 2005; Matkovi\'c \& Guzm\'an 2005; see also the historical
review in section 3.3.3 of Graham 2013). 

Continuing from Bender et al.\ (1992), Kormendy \& Bender (2012) have claimed
that the formation physics must be very different for dwarf ETGs fainter than
$M_B = -18$ mag and ordinary ETGs brighter than $M_B = -18$ mag (see also
Wirth \& Gallagher 1984).  To support this claim, they remark how these
galaxies appear to follow different relations in diagrams involving effective
half-light radii and/or effective surface brightnesses.  However, as explained
in Graham \& Guzm\'an (2003, 2004) and Graham (2005), this is because of the use of
these particular parameters produces a continuous but {\it curved} relation.
The bend at around $M_B = -18$ mag, and the different slopes seen at the low-
and high-luminosity ends, have little to do with the formation physics but are
instead a consequence of ``structural non-homology''.  That is, the bend is due
to the continually changing radial concentration of light --- which can be
quantified through the use of the S\'ersic (1963) index --- as one moves along the
ETG sequence in luminosity and mass.  See Graham (2013, 2016) for a historical
review and detailed explanation.

Characterizing the structure and kinematics of isolated,
dwarf ETGs in the low-$z$ universe will enable us to better address the origin
of dwarf galaxies.  Due to their isolation, they cannot be harassed and
morphologically transformed late-type galaxies. Therefore, the detection of a
rotating disk in these galaxies would undermine the alleged Sp/dIrr origin of
dwarf ETGs in clusters.  To reiterate that point, if field ETGs have acquired
their rotation from their original growth into an early-type galaxy,
i.e.\ Nature, then it removes the need for the transformation of late-type
disk galaxies by either the cluster environment or a neighboring massive
galaxy, i.e.\ Nurture, as the (sole) formation path for dwarf ETGs.  We have
undertaken such an observing campaign, and the results are presented in
Janz et al.\ (2017).  Here we provide a more detailed presentation for one of
these galaxies --- specifically, for the galaxy CG~611 --- than can be
afforded, or is necessary, for all of the survey galaxies.  While the
existence of just one isolated, rotating, dwarf ETG is sufficient to question
the popular formation mechanism for the rotating dwarf ETGs in clusters,
larger samples enable one to explore the extent of the over-lapping range of
physical properties of cluster and isolated dwarf ETGs.  The process of disk
growth, and disk removal, can leave galaxies with disks of varying, intermediate,
sizes.  In this paper, we present some of the hitherto overlooked implications
of this, and offer ways forward.

In Section~2 we present the data on CG~611, including its location, isolation,
distance, brightness, and effective half-light parameters
(Section~\ref{Sec2a}).  In Section~\ref{Sec2b} we present a sub-arcsecond
resolution image, a color map, and a bulge/disk/etc.\ decomposition of the
galaxy light.  In Section~\ref{Sec2c}, we present new Keck/{\tt ESI} spectra
and an investigation of the kinematic behavior 
within the galaxy's half-light radius.  A discussion of this collective data
is presented in Section~\ref{Sec_31}.  We then identify similar galaxies
located in clusters (Section~\ref{Sec_other}) and present the revelations
that this galaxy provides in regard to the formation of dwarf (and ordinary)
ETGs (Section~\ref{Sec_theory}).  Building on this, in Section~\ref{Sec_spin}, 
we stress the problematic nature of representing galaxies with
intermediate-scale disks as single points in the popular spin--ellipticity
diagram used to classify their kinematic state as either a `fast rotator' or
`slow rotator' (e.g.\ (Emsellem et al.\ 2007; Cappellari et al.\ 2007).  We
suggest a way forward that accounts for both kinematic and ellipticity radial
gradients where these galaxies transition from rotating fast where the disk
dominates, to rotating slowly at larger radii.  Finally, in
Section~\ref{Sec_Sk}, we discuss how studies are combining in quadrature the
stellar rotation and the velocity dispersion from individual galaxies
($S_{0.5} = \sqrt{0.5\, V^2_{\rm rot} + \sigma^2}$), but not yet using the
sizes of the galaxy components (disk and spheroid) associated with these
kinematic terms.  We provide a brief summary in Section~4, which includes a
list of future observations that could help to further this area of research.

\begin{figure}
\centering
\includegraphics[angle=0, width=0.48\columnwidth]{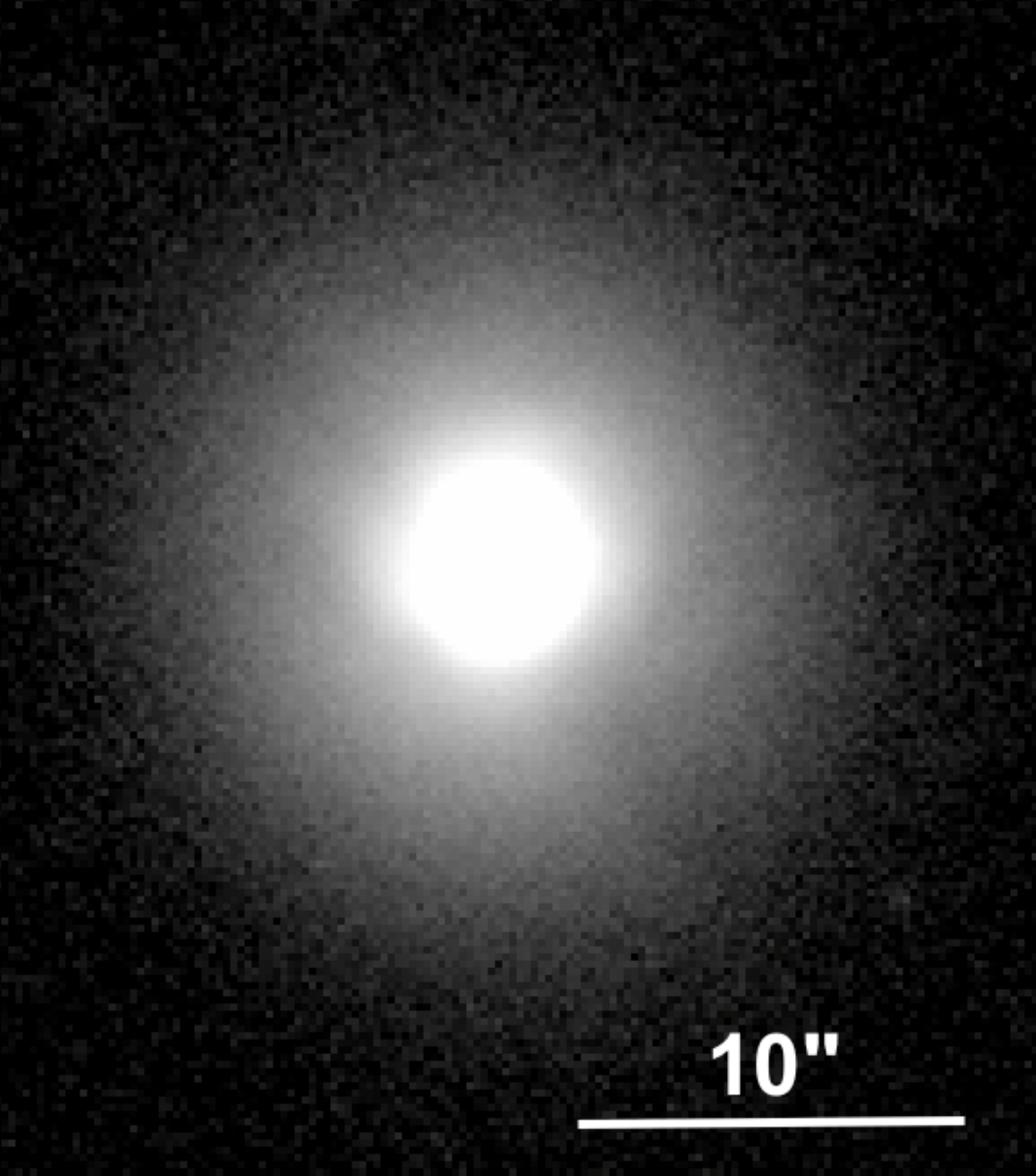}
\includegraphics[angle=0, width=0.48\columnwidth]{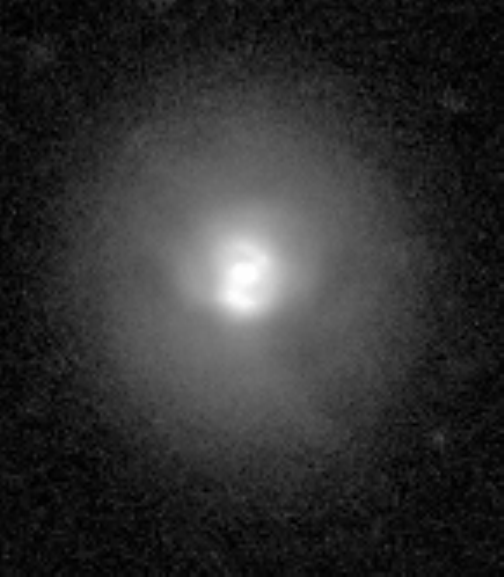}
\caption{Left panel: 
$CFHT$ $r$-band image of CG~611 ($R_{\rm e,gal} = 3\arcsec.2 = 0.71$ kpc). 
Right panel: manipulated unsharp-mask of the left panel to 
reveal the dust and inner structure. 
North is up and east is to the left in all images. 
}
\label{Fig1}
\end{figure}

\section{Data} 

Aside from being found in galaxy clusters, dwarf ETGs also exist in galaxy
groups (e.g.\ NGC~205 in the Local Group: Monaco et al.\ 2009; Saviane et
al.\ 2010) and in isolation.  Although not
commonplace, dozens of isolated dwarf ETGs are known (e.g.\ Karachentseva et
al.\ 2010; Fuse et al.\ 2012).  Building on the type of survey performed by
Paudel et al.\ (2014), one of us (S.J.P.) has recently found 46 isolated
dwarf ETGs with masses less than $5\times10^9~M_{\odot}$ and with redshifts 
$z<0.02$. 
These have been tabulated in Janz et al.\ (2017).  Rotation curves, obtained
by one of us (J.J.), are presented 
there for nine of those galaxies with $M_B \lesssim -18$ mag. 
Here we report in some detail on one of those 46 dwarf ETGs: CG~611.

\subsection{Basic data}\label{Sec2a}

LEDA~2108986 (R.A.\ = 15:03:15.57, decl.\ = +37:45:57.2, Paturel et al.\ 2003), 
previously cataloged by Sanduleak \& Pesch (1987) 
as Case Galaxy\footnote{``Case'' for Case Western 
  Reserve University.} CG~611 and listed in the 
Reference Catalog of galaxy Spectral Energy Distributions 
(RCSED\footnote{\url{http://rcsed.sai.msu.ru/catalog}}, Chilingarian et al.\ 2017) 
as object 588017627242758202, resides within the direction of the Bootes void. 

Starting from the published {\it Sloan Digital Sky Survey} (SDSS) 
Data Release 3 (DR3: Abazajian et al.\ 2005) 
heliocentric radial velocity measurement for CG~611 
of 2562$\pm$28 km s$^{-1}$, and adjusting 
for `Virgo + Great-Attractor + Shapley' infall (Mould et al.\ 2000), we 
have an expansion velocity of 3140$\pm$37 km s$^{-1}$ for CG~611. 
We assume a spatially flat universe
(i.e.\ $\Omega_{\Lambda} + \Omega_m =1$) with
$\Omega_m = 0.308\pm0.012$ and a Hubble 
constant $H_0 = 67.8\pm0.9$ km s$^{-1}$ Mpc$^{-1}$
(Planck Collaboration et al.\ 2015).  
With this cosmology, the `angular size distance' to CG~611 is 45.7 Mpc, 
which corresponds to a scale of 222 parsec per arcsecond (Wright
2006)\footnote{\url{http://www.astro.ucla.edu/~wright/CosmoCalc.html}}.
The `luminosity distance' is 46.7 Mpc, giving a (cosmological redshift
corrected) distance modulus of 33.35 mag.

CG~611 has an SDSS Petrosian $g^{\prime}$-band apparent magnitude of
15.44 mag (AB)\footnote{In the $g^{\prime}$-band, $M_{\odot,Vega} -
  M_{\odot,AB} = 0.1$.}.  After correcting for 0.052 mag of Galactic
extinction in the $g^{\prime}$-band toward CG~611 (from the Schlafly \&
Finkbeiner 2011 re-calibration of the Schlegel, Finkbeiner \& Davis 1998
infrared-based dust map as reported by
NED\footnote{\url{http://nedwww.ipac.caltech.edu}}), this corresponds to an
absolute $g^{\prime}$-band magnitude of $-17.96$ mag.  
The SDSS Petrosian $r^{\prime}$-band apparent magnitude is 14.73 mag (AB), 
and the Galactic extinction corrected $r^{\prime}$-band absolute magnitude is
$-18.66$ mag (AB). 
The average $g^{\prime} - r^{\prime}$ galaxy color of 0.7 is typical of dwarf
ETGs in the Virgo cluster (Janz \& Lisker 2009).

From a single S\'ersic fit, the published $r^{\prime}$-band, major-axis,
half-light radius $R_{\rm e,gal}$ is 3.2 arcsec, or 0.71 kpc (taken from
the NASA Sloan Atlas\footnote{\url{http://www.nsatlas.org/}}, adopting the 
sky-subtraction method of Blanton et al.\ 2011), and the mean effective surface
brightness is $\langle \mu \rangle _{\rm e,gal} = 19.22$ mag arcsec$^{-2}$.
The published S\'ersic index $n_{\rm gal}=2.3$ is typical for a galaxy of this
magnitude ($M_B \approx -18$ mag), placing it on the $L$--$n$ diagram (see
Figure~10 in Graham \& Guzm\'an 2003).

Hern\'andez-Toledo et al.\ (2010), Fuse et al.\ (2012) and Argudo-Fern\'andez
et al.\ (2015) have reported that CG~611 is an extremely isolated ETG. 
While Fuse et al.\ (2012) noted that there are no
neighbors brighter than $M_V = -16.5$ mag within a projected co-moving distance
of 2.5 Mpc and 350 km s$^{-1}$ in redshift space, they did observe that there
are two `companion' galaxies fainter than $M_V = -16.5$ mag within the void
that CG~611 resides.  These galaxies have $M_{r^{\prime}} = -15.7$ and
$-16.4$ mag ($M_* \approx 0.5$--$1.0\times10^9~M_{\odot}$) and are at
projected distances of 2.5 and 2.8 Mpc, respectively, using our cosmological
parameters given above. 

From the NASA Sloan Atlas and the Two Micron All Sky Survey (2MASS) 
redshift survey (Huchra et al.\ 2012), we too have confirmed the isolation of
CG~611.  Within a projection of 1 Mpc, there are no known galaxies with a
velocity difference of less than $\pm1000$ km s$^{-1}$ (using the
environmental search tool in NED), and there is just one dwarf galaxy (with a
heliocentric recessional velocity of 4037 km s$^{-1}$, and $M_{r^{\prime}} =
-17.5$ mag) within a projection of 2 Mpc.  The closest `neighbor' with a
velocity difference of less than $\pm1000$ km s$^{-1}$ and which is brighter
than $M_K = -23$ mag ($M_* \approx 10^{10}~M_{\odot}$) are the NGC~5899/NGC~5900
pair, at a projected distance of 3.9 Mpc away.  
The NGC 5929 Group is 5.6 degrees away on the sky, 
or 4.5 Mpc using a scale of 222 parsec per arcsecond.

\begin{figure*}
\centering
\includegraphics[angle=0, trim=0.0cm 6.0cm 0.0cm 0.0cm, height=5.5cm]{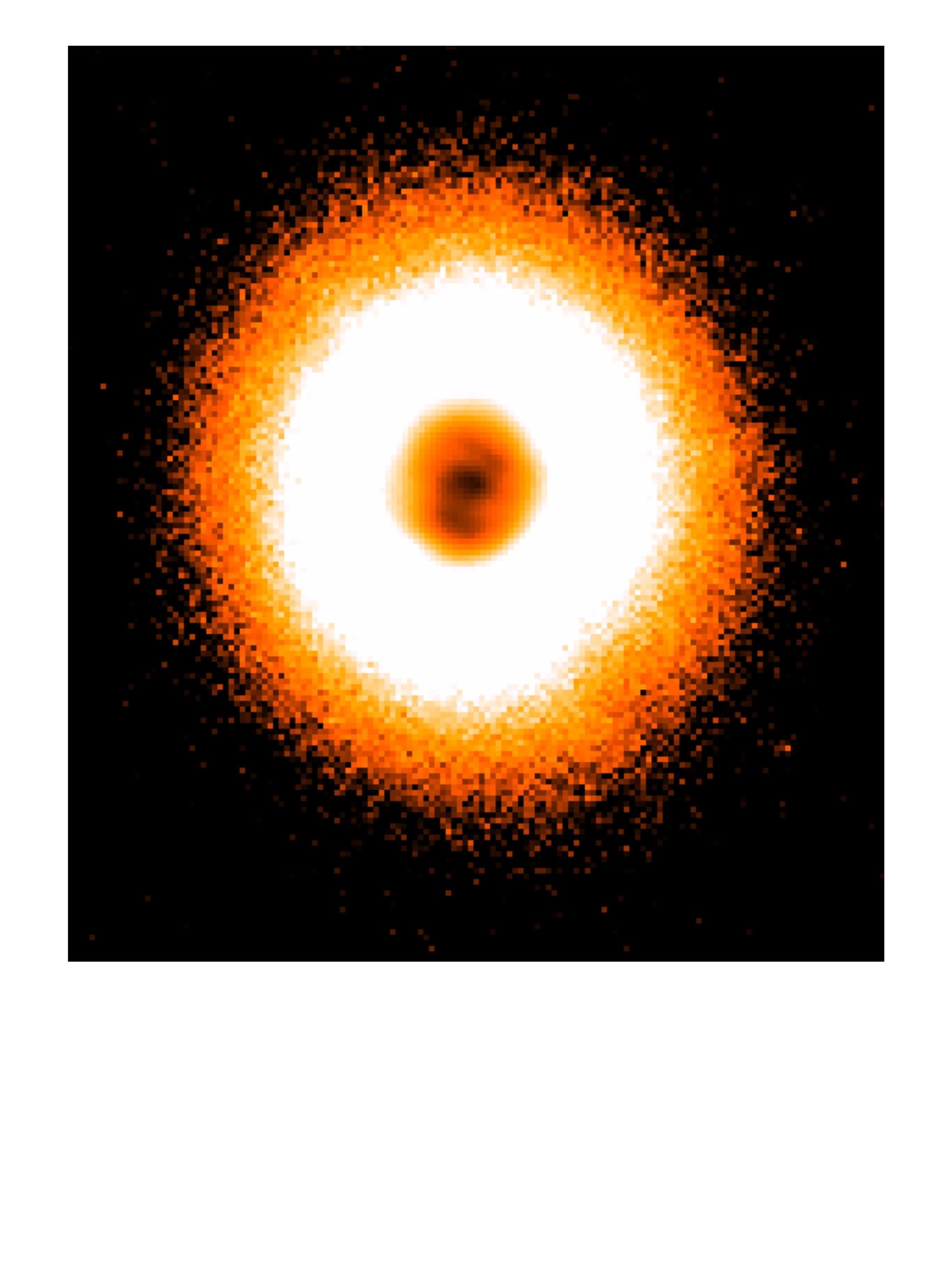}
\includegraphics[angle=0, height=5.17cm]{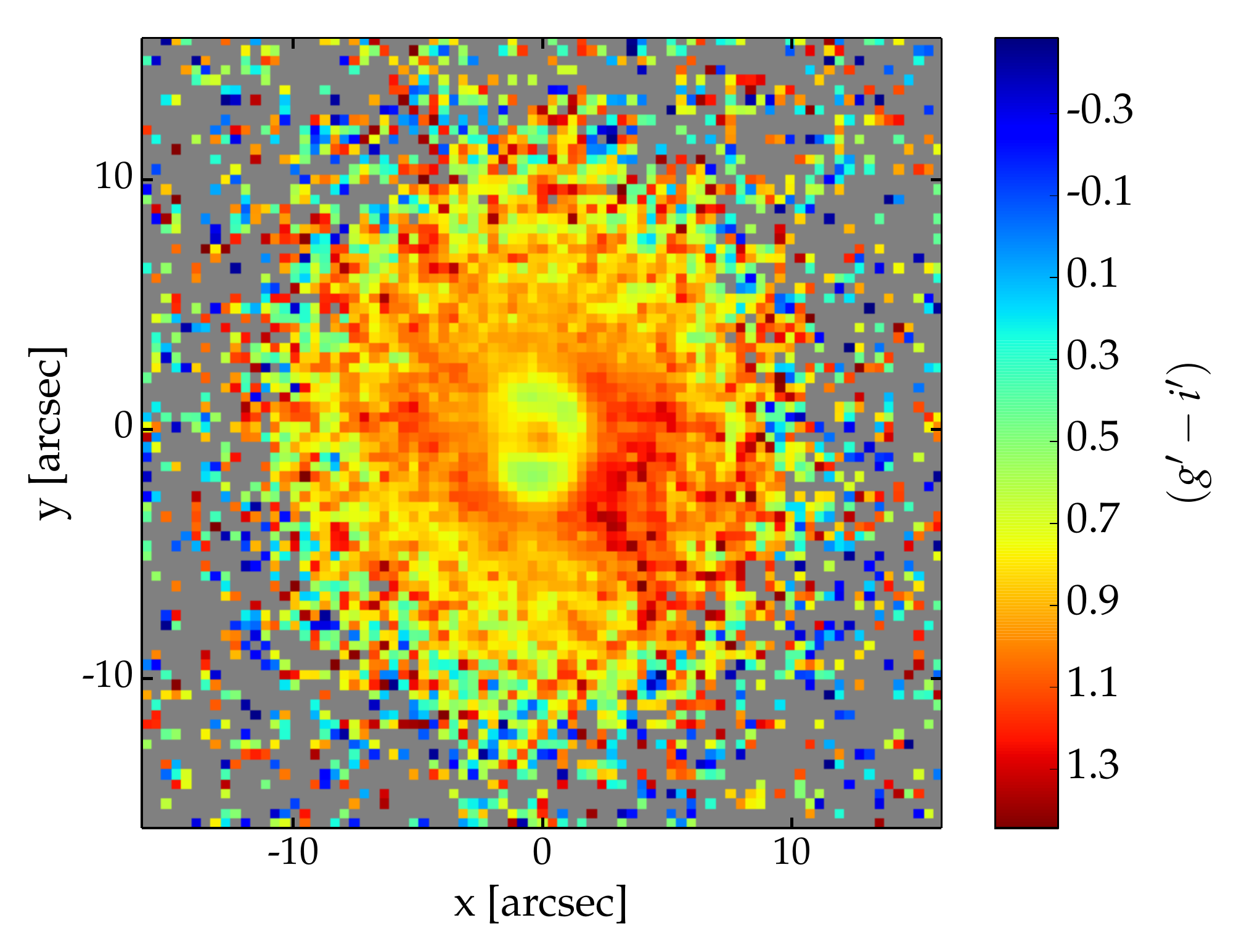} 
\caption{
Left panel: false color $CFHT$ $r$-band image to better reveal the inner bar and spiral.
Right panel: SDSS ($g^{\prime} - i^{\prime}$)-band color map of CG~611.
 Correcting for Galactic dust extinction makes the entire color map bluer by 0.025. 
}
\label{Fig2}
\end{figure*}

\subsection{Imaging Data}\label{Sec2b}

We have investigated archival {\it Canada France Hawaii Telescope} ({\it CFHT})/MegaCam
$r$-band images of the field containing CG~611 that were taken in
mid-2011 by Cheng Li (Proposal Id 11BS09).  
The full width at half maximum (FWHM) for the point spread function (PSF)
is 0.$\arcsec$8 --- measured from stars in the image. 
The galaxy is shown in 
the left panel of Figure~\ref{Fig1}, while the right panel shows a 
manipulated image to enhance the inner spiral and dust. 
If one was to strip out the dust (and gas), CG~611 would resemble other 
dwarf ETGs in clusters.  A false color image is shown in Figure~\ref{Fig2},
providing our clearest view of the nuclear spiral pattern within the much larger
elliptical stellar system. 

While we do not rule out that the main body of this ETG may be a (thick) 
disk, as in S0 galaxies, it is easy to understand from the left panel
of Figure~\ref{Fig1} why this galaxy has been classified in the literature as an ETG.
It is neither a dwarf irregular (dIrr) galaxy nor a late-type spiral galaxy in
which the spiral arms extend to large radii.  This may be a 
triaxial-shaped galaxy with an intermediate-scale disk that has grown a faint
bar with spiral arms at the ends of it.  Residing midway between E and S0
galaxies (with large-scale disk), such objects with intermediate-scale disks
were designated as ``ES'' galaxies by Liller (1966).  They have also been
referred to as ``disk ellipticals'' (Nieto et al.\ 1988), ``dEdi'' galaxies
if they are dwarfs (Lisker et al.\ 2006), and in Graham et 
al.\ (2016) the term ``ellicular'' was used to better capture the bridging
nature between elliptical and lenticular galaxies.

\subsubsection{Color map} 

Although we do not have sub-arcsecond imaging in two filters for CG~611, 
we have constructed a ($g^{\prime} - i^{\prime}$) color map using SDSS
images (Figure~\ref{Fig2}).  
The color map reveals a number of things before becoming noisy at
large radii.

The spiral arms have a blue SDSS $g^{\prime} - i^{\prime}$ color range of
0.5--0.6, commensurate with Sd--Sm type galaxies (Fukugita et al.\ 1995).  The
hot young stars that make the spiral arms shine blue likely account for the
bulk of the ultraviolet radiation coming from this galaxy.  The GALaxy
Evolution eXplorer ({\it GALEX}) All-Sky Catalog based on {\it GALEX} General
Release Six (Seibert et
al.\ 2012)\footnote{http://www.galex.caltech.edu/researcher/data.html} reports
an observed near-UV magnitude of 18.32 mag and a far-UV magnitude of 18.60 mag
(AB) for CG~611.

In stark contrast to the arms, the surrounding spheroid is red and displays 
a gradient with less red colors at larger radii --- a characteristic 
that is commonly observed in ETGs. 
Paudel et al.\ (2014) similarly report on a significant metallicity gradient
--- which is likely responsible for the color gradient --- 
in another isolated compact ETG, which drops by 0.5--0.6 dex over the inner
half-light radius ($\sim$600 pc).  We shall return to that galaxy in
Section~\ref{Sec_Disc}.  For reference, large S0 and E galaxies have an average
$g^{\prime} - i^{\prime}$ color of 1.0 and 1.2, respectively (Fukugita et
al.\ 1995), while lower mass ETGs are somewhat bluer (e.g.\ Forbes et
al.\ 2008, their Figure~1). 

For completeness, we note that the 2MASS\footnote{http://www.ipac.caltech.edu/2mass}
(Jarrett et al.\ 2000) extended objects catalog reports a total $K_s$-band
magnitude of 11.93 mag (Vega)\footnote{13.84 mag (AB).} for CG~611.  
The galaxy was also observed by $WISE$ (Wright et al.\ 2010) during the $WISE$ full
cryogenic mission. It has $m_{3.4} = 12.01$, $m_{4.6} = 12.02$,
and $m_{\rm 12} =$ 8.42 mag (Vega). 
Its red mid-infrared color [4.6]$-$[12] = 3.6 is
consistent with dust heating, and places CG~611 on the region of the [3.4]$-$[4.6] 
versus [4.6]$-$[12] $WISE$ color-color diagram occupied by spiral galaxies
(e.g.\ Jarrett et al.\ 2011). 

The innermost arcsecond is also red, suggesting the presence of a red bulge and/or
bar having an older stellar age and/or higher stellar metallicity.  
This is not an artifact due to different seeing in either image 
because the better image ($i^{\prime}$-band) was degraded (by convolution with a
Gaussian) to match the seeing of the less-well-resolved image
($g^{\prime}$-band). However, the presence of dust can act to redden the
image. 

In the color map, the dark red patch to the southwest of the spiral arms
reaches a $g^{\prime} - i^{\prime}$ color of 1.3 to 1.4 and is likely due to
dust.  There is also evidence of a dark patch at this location in the $r$-band
image, and the isophotes correspondingly pinch inward here.  Two faint,
parallel, dust lanes can also just be made out on the northeast quadrant.  If
not for the dust and the blue spiral arms, the $g^{\prime} - i^{\prime}$
color over the inner $\approx 2 R_{\rm e,gal}$ is around 0.9, typical for an
ETG of CG~611's brightness.  This presence of the dust is interesting because
it implies gas, and this is at a radius beyond the spiral arms; this is perhaps
suggestive that the intermediate-scale disk and spiral may yet grow further.
Obvious, but worth stating, is that if this disk grows sufficiently, it may
transition to a large-scale disk and the galaxy might accordingly transition
from an ES ``ellicular'' galaxy to an S0 lenticular galaxy or a spiral galaxy.
On the other hand, if the galaxy loses its gas and dust, then it will
immediately resemble dwarf ETGs in clusters.  Therefore, in this manner,
CG~611 appears similar to the ‘‘transition-type dwarf’’ galaxies mentioned by
Grebel et al.\ (2003).

\begin{figure}
\centering
\includegraphics[angle=0, width=\columnwidth]{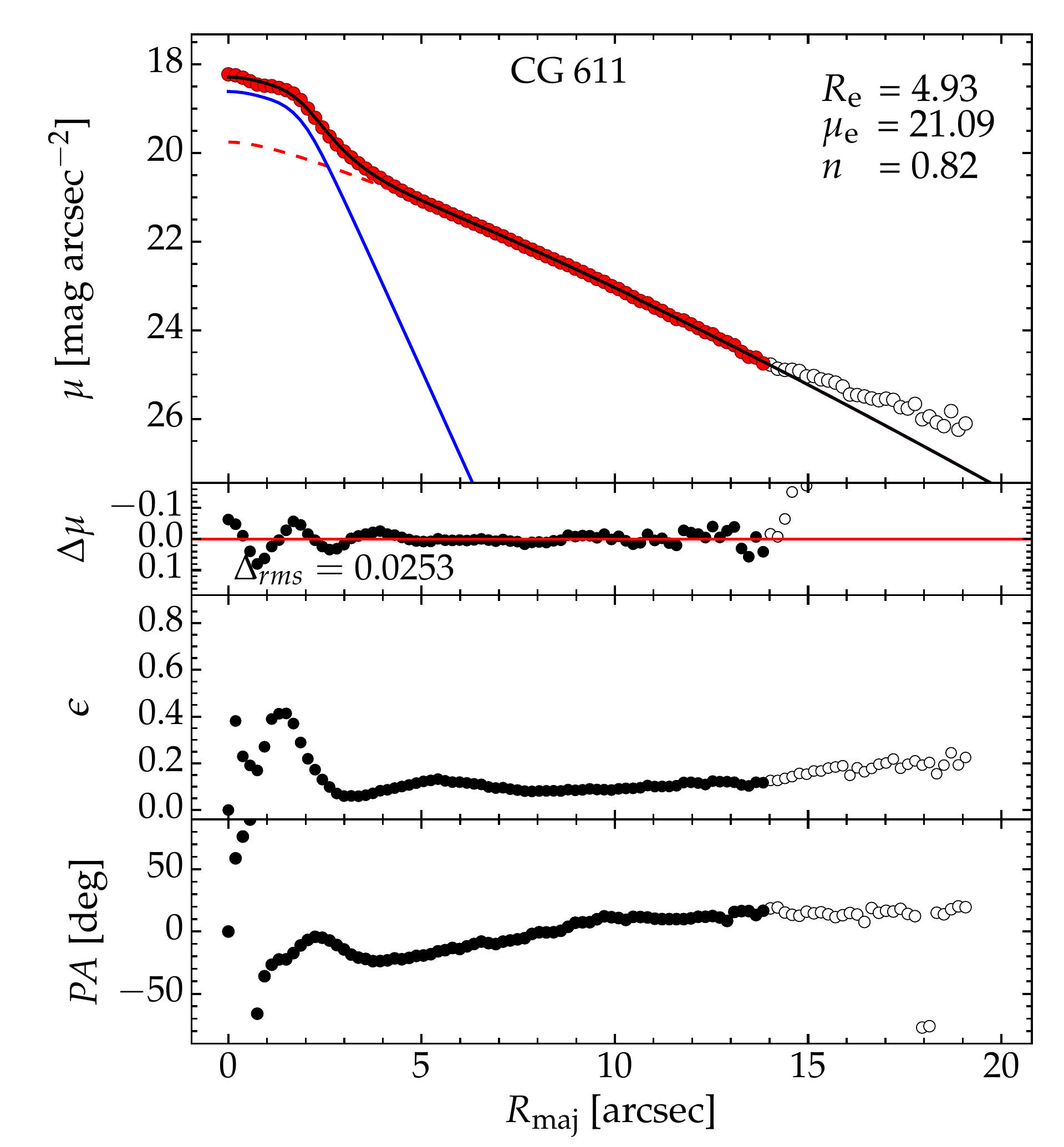}
\caption{Upper panel: the observed (i.e.\ uncorrected for Galactic Extinction, 
  redshift dimming, and $K$-correction) major-axis $CFHT$ 
  $r$-band surface brightness profile of CG~611.  
  The S\'ersic (red) plus inner truncated disk (blue) model components have
  been convolved with the PSF.  The best-fitting S\'ersic parameters (prior to
  convolution with the PSF) are inset in the figure. The quantity 
   $\Delta_{rms} = \sqrt{ \sum_{i=1,N} ({\rm data}_i-{\rm model}_i)^2/(N-\nu)}$,
where $N$ is the number of data points used (filled circles) 
and $\nu$ is the number of model parameters involved in the fit. 
  The middle panel shows the isophotal ellipticity profile 
  $\epsilon=1-b/a$, where $b/a$ is the ratio of the minor-to-major axis of
  the isophotes, while 
  the lower panel shows the position angle of the isophote's major axis. The
  open circles beyond $\sim$14$\arcsec$ have been excluded from the fit, but
  an additional outer component could be added. 
  See section~\ref{Sec_pro} for more details.  For reference, the major-axis
  half-light radius reported by SDSS for this galaxy is $3\arcsec.2$. 
 }
\label{Fig3}
\end{figure}

\subsubsection{Light profile}\label{Sec_pro}

After careful sky-subtraction and masking of interlopers, 
we used the new {\tt IRAF} task {\sc ISOFIT} (Ciambur 2015) to quantify
the shape of the isophotes and extract the surface brightness profile and associated
isophotal profiles (e.g.\ ellipticity, position angle, $B_4$, etc.) 
that describe the 2D distribution of the light.  
The surface brightness profile was then modeled 
with galactic components using the {\sc Profiler} software (Ciambur 2017).

We conducted our investigation from two stand points.  First, we considered
CG~611 to be dominated by a triaxial spheroid beyond $\approx4\arcsec$, using
the S\'ersic model to match the bulk of the galaxy light. 
Second, we assumed that this light instead comes from a thick disk, using 
an exponential model instead.

\begin{figure}
\centering
\includegraphics[angle=0, width=\columnwidth]{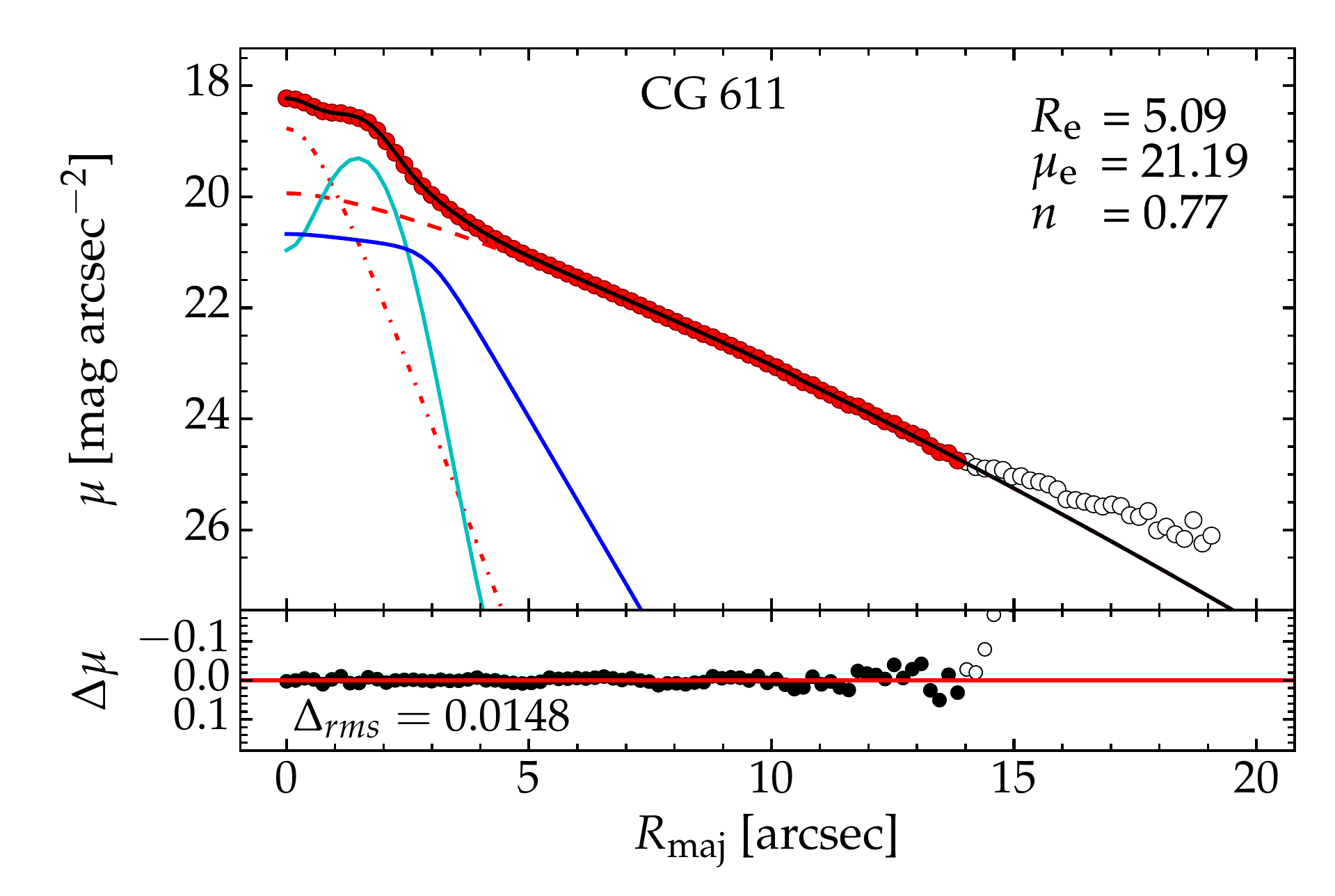}
\caption{Building on Figure~\ref{Fig3}, 
  a Gaussian function (cyan) has been added for the spiral arms/ring and a
  S\'ersic function (dash-dotted red curve) for the inner barlens. 
The S\'ersic parameters inset in this panel 
pertain to the outermost dashed red profile. 
  }
\label{Fig4}
\end{figure}

Along the major-axis, from $\approx4\arcsec$ to $14\arcsec$, the curvature of
the light profile is described well with a S\'ersic index equal to
0.82$\pm$0.05 (see Figure~\ref{Fig3}).  It is perhaps worth noting the
obvious, that this is different to the exponential value of 1 used for disks
(Patterson 1940; de Vaucouleurs 1957; Freeman 1970; Elmegreen \& Struck 2016,
and references therein).  The half-light radius of this component is 5.1$\pm$0.5
arcsec, equal to 1.1 kpc, which is typical for dwarf ETGs 
(e.g.\ Graham et al.\ 2006, their Figure~1b).  Beyond 14$\arcsec$, the surface
brightness profile starts to fall away less sharply and the ellipticity
profile slowly rises.  This behavior is also present in the SDSS images and
remains when the small 1-sigma uncertainty on the sky-background flux is subtracted
from the light-profile, which has itself already had the sky-background flux
subtracted from it.  While we refrain here from adding an additional component,
such as an exponential, we will return to this feature in the Discussion section. 

At small radii, a truncated disk model (van der Kruit 1987; Pohlen et
al.\ 2004) has been used to collectively account for the inner excess of light
where a bulge-like feature (possibly a barlens: Laurikainen et al.\ 2011), 
a bar, and a spiral pattern reside.   
The truncation of this disk may be more dramatic than shown in
Figure~\ref{Fig3}, but a sudden `cliff-like' truncation produced similar
overall results while producing an artificial local feature in the residuals.
A non-truncated disk model did not work.

Rather than leave the decomposition at this stage, we went on to account for
the spiral arms --- which look a little like lopsided ansae or a partial ring
--- by using a Gaussian function for the associated bump in the light profile.
The presence of these spiral arms shows up clearly in the ellipticity
profile (see the middle panel of Figure~\ref{Fig3}), as does the 
additional bulge-like component, which
we modeled with a S\'ersic function.  As noted, this bulge-like component may be
a barlens, and the mismatch in position angle between it and the bar (see the
lower panel of Figure~\ref{Fig3}) is a common feature of bars and their inner
lenses.  The result of fitting four
components (spiral arms, barlens, truncated intermediate-scale disk, and
large-scale S\'ersic component) 
can be seen in
Figure~\ref{Fig4}.  The faint bar has effectively been subsumed into the truncated
disk and inner bulge/barlens.

\begin{figure*}
\centering
\includegraphics[trim=1.0cm 6.9cm 1.0cm 6.7cm, width=\textwidth]{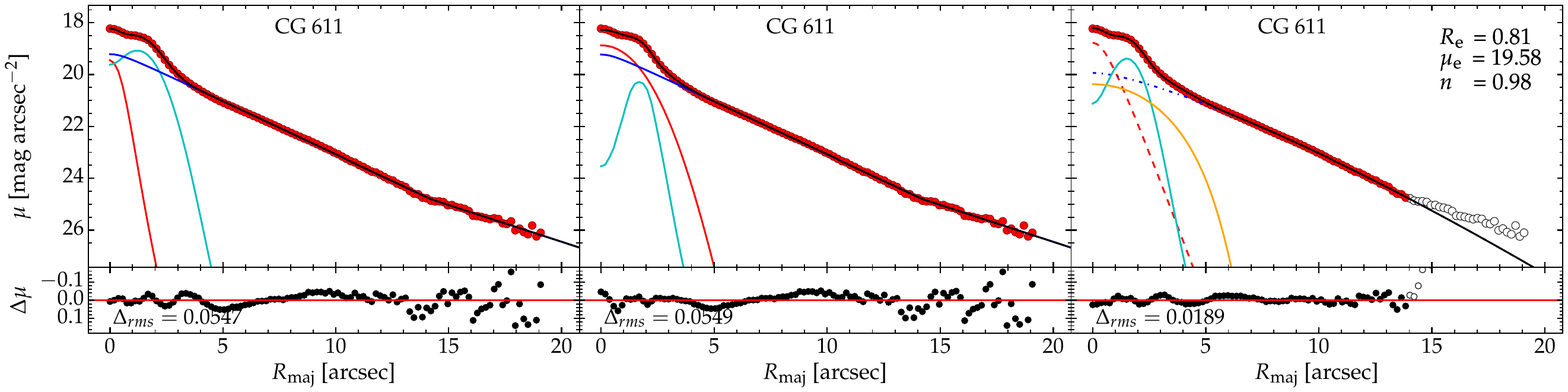}
\caption{Left panel: 
an anti-truncated exponential disk model (blue curve: with transition at
$\sim$14$\arcsec$ and inner/outer scalelength equal to $2\arcsec.7\pm0.1 / 3\arcsec.8\pm0.4$), plus 
 a Gaussian function (cyan) for the spiral arms that form a partial ring, and a
  S\'ersic function (red) for the inner bulge-like feature that may be a barlens. 
Middle panel:  an alternate fit reveals a degeneracy and thus uncertainty
with the previous decomposition whose innermost component (S\'ersic function) was not bright
enough to explain the central peak in the ellipticity profile (Figure~\ref{Fig3}). 
Right panel: including a Ferrers bar (orange) with the previous components did not
help until we truncated the data at 14$\arcsec$ 
and replaced the anti-truncated disk with the S\'ersic disk shown 
here by the dotted-dashed blue curve. The S\'ersic parameters inset in this panel 
pertain to the innermost dashed red profile.  Due to the differing radial
extent of data, the value of $\Delta_{rms}$ in this panel should 
be compared to that in Figure~\ref{Fig4}, rather than that in the left and middle
panels shown here. 
}
\label{Fig5}
\end{figure*}

The combined apparent magnitude of the above four components, when
independently fit to the inner 13$\arcsec$ of the geometrical-mean axis
(roughly 14$\arcsec$ on the major axis), and 
integrated to infinity, is 14.95 mag.  This magnitude is prior to any of the
standard corrections\footnote{Galactic extinction, redshift dimming,
  $K$-correction, and dust-inclination correction.}.  As noted previously, the
similarly uncorrected SDSS Petrosian $r^{\prime}$-band apparent magnitude is
14.73 mag (AB).  The reason that this is slightly (0.22 mag) brighter is most
likely because of the excess light beyond $\sim$14$\arcsec$ 
that we have not accounted for with our model.  The flux 
ratio of the main S\'ersic component to all four components is 0.81.

We next explored the possibility that CG~611 is comprised of a large-scale
disk rather than a spheroid.  If it is, then this would remove the need for the
inner disk shown in Figure~\ref{Fig4} because the spiral arms, bar, and barlens
could instead be features of the large-scale disk.  Such a fit is somewhat 
akin to that used for blue compact dwarf (BCD) galaxies (e.g.\ 
Sandage \& Binggeli 1984; Mrk~36 in Cair\'os et 
al.\ 2001a, or Mrk~370 in Cair\'os et al.\ 2002), 
although the color maps of BCDs (e.g.\ Cair\'os et 
al.\ 2001b) reveal a tendency for irregular off-center star-formation,
unlike that seen in CG~611 (Figure~\ref{Fig2}).\footnote{We are not 
  ruling out potential evolutionary connections with BCDs (e.g.\ Caon et
  al.\ 2005), but rather remarking that CG~611 does not presently look like a
  BCD.} 
Given the existence of
anti-truncated disks, whose surface brightness profiles display an upward bend
at large radii (e.g.\ Erwin, Beckman \& Pohlen 2005), we have employed such a
double exponential profile to represent the supposed disk.  However, we could
have used a single exponential function and continued to truncate the data at
14 arcsec, as done in Figures~\ref{Fig3} 
and {Fig4}.  

The left and middle panels of Figure~\ref{Fig5} reveal a
degeneracy that arose between the two inner components: the barlens and the
spiral arms.  Given the two peaks
over the inner ($< 3\arcsec$) ellipticity profile, neither of these two fits
appear 
satisfactory.  We attempted adding an additional bar component, but without
success.  Potentially, the dust in this galaxy may have altered the image to
yield a non-exponential disk out to 14$\arcsec$.  Allowing for this
possibility by using the S\'ersic function to represent the disk within
14$\arcsec$, we still encountered this degeneracy with our three component 
fit (barlens, spiral arms, S\'ersic-disk).  
However, we were able to overcome this when we subsequently included a Ferrers
bar (Ferrers 1877; Sellwood
\& Wilkinson 1993), as seen in the right-hand panel of Figure~\ref{Fig5}.  This is basically
the solution obtained in Figure~\ref{Fig4}, but now interpreted differently.
The radial extent of the Ferrers bar beyond the Gaussian function may indicate
that the spiral arms are not fully captured by the Gaussian function.  Use of
an anti-truncated S\'ersic disk --- whose outer extent could be exponential
in nature --- would capture the full radial extent of the data.  However, as 
far as we are aware, implementation of such a function has never been 
performed, and it is not necessary for our purposes. 

At this point, we conclude that the main body of CG~611 is probably a triaxial
spheroid but may be a thick disk without spiral structure.  In the Discussion
section, we compare CG~611 with other galaxies to see what insight we may
gain.

\subsection{Spectroscopic Data}\label{Sec2c}

The Echellette Spectrograph and Imager ({\tt ESI}: Sheinis et al.\ 2000)
on Keck II was used on the 
10th evening of January 2016 as a part of Project W138E (PI: Graham), and on the 
13th evening of March 2016 as a part of Project W028E (PI: Graham).  
To achieve our desired spectral resolution of $\sim$20 km s$^{-1}$ for the velocity
dispersion, we used  
the 0$\arcsec$.75 wide (and 20$\arcsec$ long) slit which gives an 
$FWHM$ spectral resolution
of 55.9 km s$^{-1}$.  Therefore, with a good ratio of signal-to-noise 
($S/N$) we are able to measure velocity dispersion $\sigma$ down to 
$FWHM/2.35 \approx 24$ km s$^{-1}$ (see Janz et al.\ 2017, and their Figure~5, 
for further details).  

We oriented the slit along the major-axis of the inner isophotes
(P.A.=158$^{\circ}$ east-of-north).  The combined exposures, under
0$\arcsec$.8 seeing, totaled 75 minutes.  The wide wavelength coverage of {\tt
  ESI}, from $\sim$4000 \AA{} to $\sim$10,000 \AA{} provides several spectral
features to extract the velocity dispersion, including
the calcium triplet (CaT), Mg, Fe, and H$\alpha$ lines.  To obtain the
kinematical measurements of the stars, we averaged the values (weighted with
the inverse variance squared) obtained from fits to the Mg$b$ and Ca {\sc II}
triplet regions. The $S/N$ ratio is $>$30 per Angstrom in the center for both
spectral regions and around $\sim$6-7 for the largest radii.  Radial
velocities and velocity dispersions were measured using 
the penalized PiXel Fitting (pPXF) software by
Cappellari \& Emsellem (2004) and suitably matched template stars.

\subsubsection{Kinematics}

\begin{figure}
\centering
\includegraphics[trim=1.7cm 1.7cm 1.7cm 1.7cm, width=1.0\columnwidth]{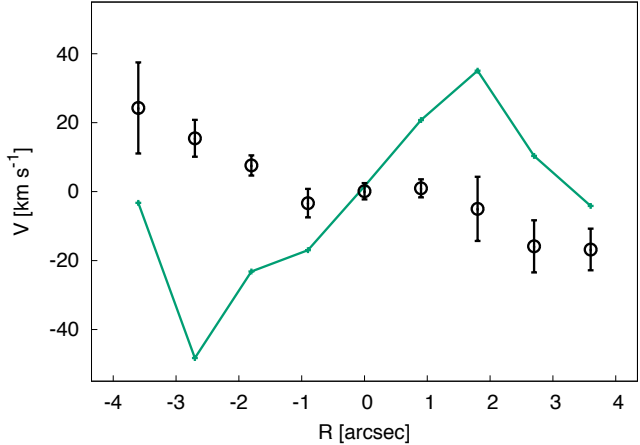}
\caption{ 
Observed rotation curve of the stars (open symbols) rises to $\approx$18 km
s$^{-1}$ by 1 $R_{\rm e,gal}$ ($=3\arcsec .2$, $= 0.71$ kpc).
The inner three data points additionally display 
evidence for a kinematically distinct core (KDC) in the form of mild counter
rotation. The rotation curve of the gas (green curve) also 
counter rotates with respect to the luminosity-weighted stellar rotation 
within $\approx$1.3 $R_{\rm e,gal}$. 
Positive velocities mean a receding motion, and 
negative radii correspond to the S/SE direction (the lower right in Figure~\ref{Fig1}).
}
\label{Fig6}
\end{figure}

We measure a heliocentric recessional velocity of 2546$\pm$4 km s$^{-1}$ from
the stellar absorption lines, in good agreement with the SDSS value of
$2562\pm28$ km s$^{-1}$.  We measure a central stellar
velocity dispersion $\sigma_{\rm cen} = 37 \pm 2$ km s$^{-1}$ (see Janz et
al.\ 2017 for further information). 
This value is greater than the spectral resolution limit of $\sim$24 km
s$^{-1}$ noted in the previous section, and is therefore a robust measurement
rather than a lower limit.

\begin{figure}
\centering
\includegraphics[angle=0, width=1.0\columnwidth]{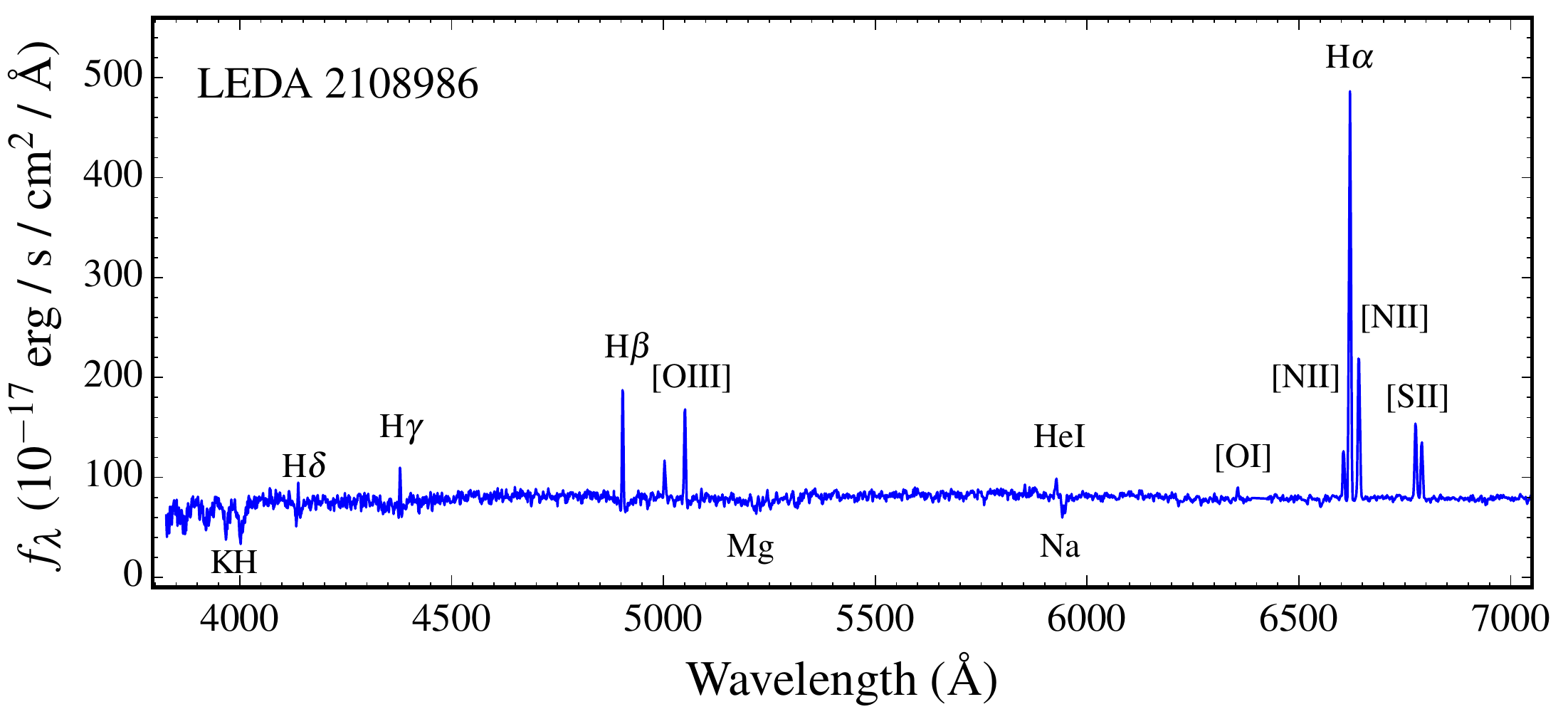}
\caption{ 
SDSS spectrum of LEDA~2108986 (CG~611) with notable emission and absorption lines marked. 
}
\label{Fig7}
\end{figure}

Figure~\ref{Fig6} reveals that 
the galaxy's stars rotate fairly symmetrically over the inner 
$3\arcsec .6$ (0.80 kpc, 1.1 $R_{\rm e,gal}$). 
The positive stellar rotation (the left half of Figure~\ref{Fig6})
corresponds to the southern spiral arm in Figure~\ref{Fig1}. 
As can be seen in Figure~\ref{Fig3}, 
this radial extent of 4$\arcsec$.2 
also happens to roughly correspond with the region where
the disk (including the spiral arms) contributes to the light profile
above that of the main S\'ersic component. 
The stellar rotation at 1 $R_{\rm e,gal}$ is $\approx$18 km
s$^{-1}$ without correcting for the inclination of the disk.  
Based on the minor-to-major axis ratio of $\sim$0.90 at 1 $R_{\rm e,gal}$ (see
Figure~\ref{Fig3}), this would correspond to an intrinsically round 
disk's inclination of $\sim$26 degrees; implying that the full rotational speed is
$\approx$41 km s$^{-1}$ in the plane of the disk.

The galaxy is additionally known to have several emission lines (see
Figure~\ref{Fig7}).  Fitting these independently in our Keck/{\tt ESI} data
with the pPXF software reveals the ionized gas dynamics, which is
counter-rotating with respect to (most of) the stars within $\approx$1 $R_{\rm
  e,gal}$ (Figure~\ref{Fig6}).  Such retrograde motion in barred galaxies has
long been observed (e.g.\ Galletta 1987; Corsini 2014).  With a peak gas rotation
of $-48$ km s$^{-1}$ at $-2\arcsec .7$ ($-0.84 R_{\rm e}$) and $+32$ km
s$^{-1}$ at $+1.\arcsec 8$ ($+0.6 R_{\rm e}$), this gives a $V_{\rm
  gas}/\sigma_{\rm cen}$ ratio of around $-1$.  These peaks roughly occur at
the peak density of the stellar spiral arms.

There is additionally tentative evidence for a kinematically distinct component (KDC) in
the form of a counter-rotating stellar core involving the inner $\sim$0.8 to 
$\sim$1.6 arcseconds (0.25 to 0.5 $R_{\rm e,gal}$). 
This roughly corresponds with the radial extent of the barlens and the
faint bar, respectively.  
The amplitude of the counter rotation is marginal, at just a few km
s$^{-1}$, although this feature is of course diluted by the signal coming from
all the other stars along the same lines of sight through the galaxy.  
Amplitudes of counter rotation in other dwarf ETGs are also 
observed to be low, at 
around $\sim$5 km s$^{-1}$ (Geha et al.\ 2003).  It is possible that some new
stars may have formed out of the accreted gas, which could explain this slight
KDC.  This would imply that yet another component should be added to the
decomposition of the light profile in Figures~\ref{Fig4} and \ref{Fig5}.  However,
doing so would result in too much degeneracy and is therefore left for
when higher spatial resolution imaging data becomes available.

\section{Discussion}\label{Sec_Disc}

\subsection{CG~611}\label{Sec_31}

Given that CG~611 lives in a void, it is obviously not a late-type
spiral galaxy that has been morphologically transformed by a cluster
environment or a nearby massive galaxy --- the argument often advocated for
the creation of rotating dwarf early-type galaxies.  CG~611 is also not
hurtling through space at high speed, perhaps ejected from some distant galaxy
cluster 
(e.g.\ Gill et al.\ 2005; Chilingarian \& Zolotukhin 2015), because it is too
far from any cluster, and if it were moving at high speed, then it 
would not have accreted its counter rotating gas disk. 

As noted in Section~\ref{Sec_pro}, 
CG~611 possibly consists of an intermediate-scale 
stellar disk embedded within a much larger spheroid (see Savorgnan \& Graham
2016b for other examples of such ES galaxies).  Ann, Seo \& Ha (2015) reclassified
CG~611 from E/S0 to SA0/a; most likely because they detected the faint bar and the
weak spiral structure from their visual inspection.  However, relative to the
S0 galaxy type, they shifted CG~611 slightly toward the late-type
classification ``Sa''.  It might be appropriate to shift it slightly toward the elliptical
galaxy classification by recognizing it as an ES galaxy (see Figure~7 in Graham
et al.\ 2016) in which CG~611 does not have a large-scale disk 
that dominates at large radii.  That is, CG~611 may not be a disk galaxy
with an inner bulge component, but instead an elliptical galaxy with an 
inner disk component. 

Koleva et al.\ (2009, and references therein) report that dwarf ETGs in
clusters and groups are, in general, dominated (in stellar mass) by old stars,
which are ``likely coeval with that of massive ellipticals or bulges, but the
star formation efficiency is lower''.  Due to additional intermediate-age (1-5
Gyr) populations and tails of ongoing star formation, their inner regions have
(single-stellar-population)-equivalent ages of 1 to 6 Gyr.  
Ry\'s et al.\ (2015) also report that star formation within the last 5 Gyr is
common in dwarf ETGs. They have interpreted this as support for the formation
scenario in which tidal harassment drives gas (not stripped due to the ram
pressure of the cluster/group hot X-ray gas) inward and induces a 
star formation episode. However, CG~611 is evidence that this scenario need
not necessarily be true, or rather is not the only mechanism to account for
intermediate-age cores in dwarf ETGs.  This is because the external accretion of gas
may have, in part, built its young nuclear region.  The observation that a lower
fraction of ordinary ETGs in clusters, than compared to isolated ETGs, are
blue and star-forming (Lacerna et al.\ 2016, see also Collobert et al.\ 2006) 
reflects this and reveals a greater abundance of 
cold molecular gas in the field galaxies (see also Moorman et al.\ 2016). 

Accretion of gas in CG~611 is apparent from the counter
rotation of the gas relative to most of the stars, and from the
non-symmetrical nature of the gas rotation curve about the photometric center
of the galaxy (Figure~\ref{Fig6}).  
Ionized gas kinematics from future IFU data may help to establish when the gas
in CG~611 was accreted; for example, offset rotation implies that it was
acquired recently. We do not know the kinematic position angle of the ionized
gas. 
An offset from the stars by $\sim$0 or $\sim$180 degrees 
implies dynamical equilibrium with the stars, and that the gas 
could have been in the galaxy for some time, while offsets greater than 30 degrees 
would suggest more recent accretion events.  
It would also be interesting to know if there 
is a bigger neutral hydrogen cloud around CG~611, as we only see the
ionized gas that has been photoionized by the hot blue stars in the spiral
arms.  
Curiously, the small spiral arms in CG~611 bare an uncanny resemblance to the
nuclear spiral structure established in circumbinary disks around binary stars
with co-rotating orbits (Diego Pinto et al., in preparation).  Rather than two
planets, the two over-densities at the ends of the bar give rise to the trailing
spirals.

CG~611 has marginal evidence for a kinematically distinct {\it
  stellar} core (KDC) of $\sim$220--350 pc in diameter.  KDCs have been
observed in dwarf ETGs before, and Toloba et al.\ 2014 report on a 140 pc and
330 pc (in radius) KDC in the Virgo cluster galaxies VCC~1183 and VCC~1453.
Penny et al.\ (2016, see their Figure~8) report on two dwarf ETGs with 1-1.5
kpc KDCs.  Morphological peculiarities are common to isolated ETGs (Reda et
al.\ 2004) and are therefore not an identifying 
signature of galaxy harassment or tidal stirring in cluster ETGs. 

Far from being unusual, the majority of isolated lenticular galaxies contain
ionized gas disks, and half of those disks rotate in the opposite sense to the
stellar disk (Katkov et al.\ 2014, 2015).  As Katkov et al.\ (2015) remark,
this is expected if all of the gas is acquired by accretion of gas-rich
satellites, or perhaps filaments of cold gas.  For similar results, see 
Davis et al.\ (2011), Alatalo et al.\ (2013), Lagos et al.\ (2015), and also Jin et
al.\ (2016) in regard to the MaNGA (Mapping Nearby Galaxies at Apache Point
Observatory, Bundy et al.\ 2015) integral field spectroscopic survey.  The
Sydney-AAO Multi-object Integral field spectrograph (SAMI) Galaxy Survey
(Croom et al.\ 2012; Bryant et al.\ 2015) has also found that $>40$\% of
$z<0.1$ ETGs outside of clusters have evidence for recently accreted gas
(J.\ Bryant et al.\ in preparation).  The rotation of this ionized gas displays
misalignment with the stars motion and counter rotation.  Such galaxy
growth via disk growth likely accounts for the fate of many compact massive galaxies
observed at high redshift; this is necessary to explain the abundance of
compact massive spheroids in today's early-type galaxies (Graham 2013; Graham
et al.\ 2015) and likely occurs quickly (see Papovich et al.\ 2016) so that
their disks are suitably old by today.\footnote{Less massive ES and S0
  galaxies have younger disks than their more massive brethren.}  Ordinary
(i.e.\ non-dwarf) lenticular galaxies have long been known to contain old
mass-weighted spheroids and younger disks (see section~4.3.3 of Graham 2013,
and references therein) while the most massive ETGs --- the giant elliptical
galaxies --- likely formed from major dry merger events in which their partially
depleted cores were created through the damage of the ensuing supermassive
black hole binary (Begelman et al.\ 1980; Merritt \& Milosavljevi{\'c} 2005;
Merritt 2006; Dullo \& Graham 2014).

Finally, we comment on the previous observation that CG~611 follows the
correlation between galaxy luminosity and galaxy S\'ersic index for ETGs.  Had
CG~611 not grown its inner components, that is, if its surface brightness
profile within the inner 14$\arcsec$ ($\sim$3 kpc) was described by just the
S\'ersic component shown in Figure~\ref{Fig3}, then it would appear as a
low-$n$ (high-$L$) outlier in the $L$--$n$ diagram.  CG~611 would shift from
$M_{g^{\prime}}=-17.96$ mag and $n = 2.3$ (the single S\'ersic fit from SDSS)
to $M_{g^{\prime}}=-17.7$ mag and $n = 0.8$, positioning it rather close to
VCC~9 (IC~3019) in Figure~10 of Graham \& Guzm\'an (2003). This therefore
offers a potential explanation for why VCC~9 is an outlier in this diagram ---
it may not have accreted sufficient material at its core to increase its
galaxy S\'ersic index.  CG~611 additionally raises the question as to whether
or not past bulge/disk decompositions have got the bulge and disk components
around the right way.  For example, the interesting compact elliptical galaxy
LEDA~3090323, with an AGN and suspected $\sim$$2\times10^6~M_{\odot}$ black
hole (Paudel et al.\ 2016a), may be worth exploring further in this regard.

\subsection{Other galaxies}\label{Sec_other}

It is relevant to ask if CG~611 is unique in appearance, or instead similar to
other known galaxies.  Disk features and faint spiral structures have indeed
been reported in dwarf ETGs before (Jerjen et al.\ 2000; Barazza et al.\ 2002;
Graham \& Guzm\'an 2003; Graham et al.\ 2003; Lisker et al.\ 2006; Lisker \&
Fuchs 2009; Janz et al.\ 2012; Penny et al.\ 2014).  Perhaps of particular relevance
is the ``Emerald Cut''
galaxy (LEDA 074886: Graham et al.\ 2012), a dwarf ETG with an
intermediate-scale disk seen edge-on.  In that galaxy, the disk rotates at
33$\pm$10 km s$^{-1}$, compared to our estimated edge-on rotation of
$\approx$41 km s$^{-1}$ for CG~611.  LEDA~074886, and in particular its
isophotes, illustrates well how the disk is confined to within the galaxy.
CG~611 {\it may} be something of a face-on analog to LEDA~074886.  It can,
however, be challenging to obtain kinematic data at large radii to reveal the
kinematic decline of disks in dwarf galaxies.  Although, this has been
achieved for non-dwarf ES galaxies with intermediate-scale disks,
e.g.\ NGC~3115 and NGC~3377 (Arnold et al.\ 2011, 2014).

As to other dwarf ETGs displaying more face-on disks than LEDA~074886, 
the inner spiral structure in CG~611 is very similar to the Virgo 
cluster dwarf ETGs VCC~216 and VCC~490 (see Figure~2 from
Lisker et al.\ 2009 which displays these galaxies and their unsharp masks).
NGC~4150, a member of a sparse group, additionally resembles CG~611.  We
have therefore discovered that the cluster and group environment are {\it not} required
for creating these dwarf ETGs.  The residual image (data minus
symmetric model) of the ``compact elliptical'' galaxy J094729.24+141245.3 in
Huxor et al.\ (2013) also appears to show tentative evidence for an inner
spiral structure similar to that seen in CG~611.  While those authors argued that
there was no underlying disk in J094729.24+141245.3, if this spiral pattern is
confirmed --- with say {\it Hubble Space Telescope} imaging --- then this
galaxy may likely also contain an intermediate-scale disk.  
In passing, we note that the compact galaxy LEDA~083546 (Paudel et al.\ 2016b) 
also possesses several similarities to CG~611.  

Below, we focus on two galaxies resembling CG~611, but that have been treated
rather differently in the literature.  This adds to the diagnostic dilemma we
encountered in Section~\ref{Sec_pro}, in which it was not obvious whether
CG~611 contains a large-scale disk or an intermediate-scale disk within a
large-scale spheroid.

\subsubsection{NGC~2983} 

The non-dwarf ETG NGC~2983 ($M_B \approx -20.5$
mag; $B-V\approx 0.9$) is remarkably similar to CG~611 in appearance, 
with the major-axis position angle of its ``barlens'' 
considerably rotated with respect to its bar 
(Laurikainen et al.\ 2011, their Figure~8; see also Bettoni et al.\ 1988). 
Like CG~611, NGC~2983 also contains short spiral arms at the ends of its bar. 
Bettoni et al.\ (1988) report on the presence of
an additional, larger ``lens'' from $\sim$20 to $\sim$50 arcseconds (see their
Figures 1c and 3)\footnote{The 10$\arcsec$ scale-bar in their Figure~1 should
 read 20$\arcsec$.}, which is evident in the Carnegie-Irvine Galaxy Survey 
(Ho et al.\ 2011) image\footnote{\url{https://cgs.obs.carnegiescience.edu}}
and is something of a broad, ring-like, annular extension of the spiral arms.  
This ``lens'' plus the outer regions of this galaxy 
are modeled with a single exponential disk by Laurikainen et al.\ (2010), who 
therefore treat this galaxy as an SB0.  Given that the axis ratio beyond 50$\arcsec$
is around 0.57 in this galaxy, it is understandable why this was done. 
However, beyond the $\sim$20$\arcsec$-long bar, the rotation curve can be seen to decline
(albeit only by 25\%) as one starts to move outward through the ``lens''.  Despite
this, Bettoni et al.\ (1988) remark/assume that the structure beyond this is
probably oblate with an intrinsic flattening of 0.25. 
This raises the question of whether CG~611 might consist of a large-scale disk
hosting a small faint bar with a central barlens and short spiral arms coming off
its ends, plus (perhaps) a larger ``lens'' or ring (see Buta \& Crocker
1991) that extends the spiral arms out to 4--5 arcseconds.  
If so, then the bar component in the right-hand panel of Figure~\ref{Fig5}
might have extended out beyond the Gaussian component --- used to represent
the spiral arms/ring --- because it was additionally capturing such a
``lens''.

\subsubsection{CGCG~036-042}

The $r$-band surface brightness profile of CG~611 (Figure~\ref{Fig3}) is
also remarkably similar to that of the isolated\footnote{See Section~2.1 of Paudel
  et al.\ (2014).}, compact ETG CGCG~036-042 (Paudel et al.\ 2014, see their
Figure~2).  However, this galaxy is not presented as hosting a large-scale
disk.   
Modeling the major-axis light profile of CGCG~036-042, Paudel et
al.\ (2014) find that the best fit has an inner S\'ersic component with
$n=1.06$, and an outer S\'ersic component with $n=0.78$ (compare our
Figure~\ref{Fig3}).  Furthermore, they 
report on an extended diffuse stellar component 
from $\sim$3 kpc out to 10 kpc.  Beyond $\sim$3 kpc, we too observe an
extended feature in the light profile of CG~611 that appears 
similar to the extended feature in CGCG~036-042.  For simplicity, we chose not
to model this outer feature in Figures~\ref{Fig3} and \ref{Fig4}, but it can be seen in
the upturn of the ellipticity profile and the flattening of the position angle
profile beyond 14$\arcsec$ ($\sim$3 kpc) in Figure~\ref{Fig3}.  
Moreover, CGCG~036-042 has $M_r = -18.21$ mag (Paudel et
al.\ 2014), comparable to the brightness of CG~611 ($M_r = -17.96$ mag).

The compact ETG CGCG~036-042 and the non-dwarf ETG NGC~2983 serve to
illustrate that other galaxies, not just dwarf ETGs, which are similar in
appearance or light profile to CG~611 exist, and that they have been modeled
in different ways.  This echoes the ambiguity seen in section~\ref{Sec_pro},
and motivates us to develop ways to resolve this issue for galaxies in the
future.  
In particular, we introduce a diagnostic diagram in 
Section~\ref{Sec_spin} to help address this.

\subsection{Implications}

\subsubsection{Theoretical insight/connections}\label{Sec_theory}

What does all this imply for the ``galaxy harassment'' model in clusters
(Moore et al.\ 1996, 1998) and the related ``tidal stirring'' model, which can
operate around massive galaxies (Mayer et al.\ 2001a, 2001b)?  Both of these
models result in late-type disk galaxies forming bars, and the remaining outer
disk being stripped off.  These two processes initially produce rather flat
``naked bars'', which under select (near face-on) viewing angles superficially
look like ellipsoidal galaxies.  Sufficient agitation can eventually turn
these flat bars into triaxial spheroidal structures, and the Sagittarius dwarf
galaxy may be a good example of this ({\L}okas et al.\ 2010).  Stott et
al.\ (2007) and D'Onofrio et
al.\ (2015) present additional evidence for such transformations. 
While we do
not deny that this process can occur, we have revealed that this is clearly
not the origin of all dwarf ETGs; an observation that may have some support 
from De Lucia et al.\ (2009) who report on a deficit of post-starburst dwarf
galaxies in two clusters. 

In passing we note that the ``tidal stirring'' model applies to the
metamorphosis of galaxy disks, and thus it is different from the tidal
interaction models advocated for the creation of low-mass compact elliptical
(cE) galaxies (King 1962; Rood 1965; Faber 1973; King \& Kiser 1973; Keenan \&
Innanen 1975; Wirth \& Gallagher 1984).  These early models suggested that a
massive companion galaxy stripped off the outer layers of a normal elliptical
galaxy to create the relatively rare class of cE galaxies.  Subsequent work
has suggested that the original galaxy was instead a lenticular galaxy, and
thus the cE galaxies are predominantly the bulges of former S0 galaxies (Bekki
et al.\ 2001; Graham 2002), or the cE galaxies in isolation never grew a
substantial disk in the first place (Graham 2013).

Although CG~611, and the other isolated dwarf ETGs presented in Janz et
al.\ (2017), undermine the ``Galaxy harassment'' and ``tidal stirring'' models
for the creation of {\it all} dwarf ETGs, those two mechanisms {\it might}
preferentially act on lower mass, late-type spiral, and dwarf irregular
galaxies, turning them into dwarf spheroidal (dSph) galaxies (Klimentowski et
al.\ 2009; Kazantzidis et al.\ 2011, 2013; Paudel \& Ree 2014) by removing
much of their disk mass.  These dSph galaxies (with $M_B \gtrsim -13$ to $-14$
mag) may then be somewhat different from the dwarf/ordinary ETG sequence at
$M_B \lesssim -13$, with many of the most luminous ETGs ($M_B \lesssim -20.5$
mag) additionally displaying a departure in the luminosity--(central surface
brightness) diagram due to their partially depleted cores (Graham \& Guzm\'an
2003).  Some of the dSph galaxies would then be more connected with low-mass
late-type, dwarf Irregular, and ``transition type'' dwarf galaxies
(e.g.\ Grebel et al.\ 2003) and likely show different behaviors in certain
scaling diagrams.  Indeed, the color--magnitude diagram for ETGs already
suggests that there is a difference at $M_B \sim -13$ to $-14$ mag
(e.g.\ Penny \& Conselice 2008).  Roediger et al.\ (2016) note that quenching
of these low-mass ETGs may have occurred in groups and smaller host halos
prior to entering the cluster environment.  Although, it should be remembered
that dwarf spheroidal (dSph: $M_B \gtrsim -13$ mag) galaxies can also evolve
in isolation (Makarova et al.\ 2016), and thus this process is not a
requirement.  Furthermore, Smith et al.\ (2015, see also Smith et al.\ 2010
and Bialas et al.\ 2015) ran a suite of numerical simulations and found that
harassment of low-mass disk galaxies was only effective if they plunged deep
into the dense core of a galaxy cluster. They found that less than one-quarter
of orbits from a cosmological simulation resulted in any stripping of disk
stars, as required in the harassment scenario.  In addition, Aguerri (2016)
concluded, based on his analysis of the $(R_{\rm e,bulge}/h_{\rm disk})$--$n$
diagram, that most dwarf ETGs in clusters are not transformed disk galaxies.

In CG~611, minor mergers may have brought in some gas and stars in a clumpy
manner, which is known to lead to the growth of disks, including counter
rotating cores and halos (Sancisi et al.\ 2008; Gonz{\'a}lez-Samaniego et
al.\ 2014; Lane et al.\ 2015;
Christensen\footnote{\url{https://www.youtube.com/watch?v=_Ssc1GsqHds}} et
al.\ 2016).  
As noted by Genel et al.\ (2010), minor mergers and smooth accretion, rather
than major mergers, dominate galaxy growth (see also Murali et al.\ 2002;
Semelin \& Combes 2005; Maller et al.\ 2006), and this supply channel also
favors the formation of disks.  Smooth gas flow onto a galaxy can occur via
spherical accretion of $\sim$10$^4$--$10^5$ K halo gas (e.g.\ White \& Frenk
1991, see their Section~7; Birnboim \& Dekel 2003) or more directly from
unshocked cold streams (e.g.\ Kere{\v s} et al.\ 2005; Dekel \& Birnboim
2006; Brooks et al.\ 2009; Dekel et al.\ 2009; Cornuault et al.\ 2016).  This
cold mode of delivery may be the most efficient of the mechanisms (Genel et
al.\ 2010; van de Voort et al.\ 2011; Lu et al.\ 2011) and Pichon et
al.\ (2011) have revealed how these cold gas streams can build a disk in a
coherent planar manner (see also Danovich et al.\ 2012, 2014; Prieto et
al.\ 2013; and Stewart et al.\ 2013).

Examining the kinematics of a larger sample of
isolated dwarf ETGs will enable one to compare their properties with cluster
dwarf ETGs and in so doing learn about their collective origin.  For example,
Paudel et al.\ (2010) have suggested that old metal poor dwarf ETGs found in
the central high-density regions of clusters may have formed along with the
cluster while the younger more metal rich dwarf ETGs (which tend to have disk
features) have fallen in at a later time.  While plausible, the assumption
that the infalling galaxies were disk galaxies that underwent a structural
transformation may need revising if isolated dwarf ETGs already have the same
structural and kinematic properties as this cluster population.  CG~611 is one
such galaxy that suggests they do, and Janz et al.\ (2017) presents a further
eight such galaxies.

A related issue is the dominance at low masses of dwarf ETGs in clusters and
the prevalence of (higher mass) Scd-Im galaxies in the field.  Although we do not
answer this question, obviously past
claims that have attempted to answer this solely in terms of harassment and
transformation of late-type galaxies need to be reconsidered, because 
dwarf ETGs similar to those in clusters 
are now known to also exist in the field.  There is 
a formation channel for dwarf ETGs that does not involve the
harassment of disk galaxies.  In the $\Lambda$ cold dark matter ($\Lambda$CDM)
model of galaxy formation, low-mass dwarf galaxies naturally form, and help to
build up bigger galaxies.  One can speculate that 
these small systems might have preferentially formed
in over-dense regions that later became galaxy groups and clusters.
That is, they may have effectively formed as ETGs rather than as late-type
galaxies that were subsequently relocated and transformed.

\subsubsection{Spin--ellipticity diagrams}\label{Sec_spin}

We have seen in Section~\ref{Sec_pro} that CG~611 may have an
intermediate-scale disk, and we have learned in section~\ref{Sec_other} that
other galaxies certainly do.  Here we propose a revision to a popular diagram,
used to characterize the kinematic behavior of ETGs as either `fast rotators'
or ``slow rotators'', by displaying additional information helpful for
identifying galaxies with intermediate-scale disks.  A key ingredient is
kinematic information extending beyond the radial extent of a galaxy's disk.
While we do not yet have this information for CG~611, it exists for other
galaxies with intermediate-scale disks and should become available for many
more galaxies through present and upcoming `integral field unit' spectroscopic
surveys such as SAMI (Croom et al.\ 2012; Fogarty et al.\ 2015).

\begin{figure*}
\centering
\includegraphics[angle=0, trim=1.0cm 6.7cm 1.0cm 7.3cm, width=1.0\textwidth]{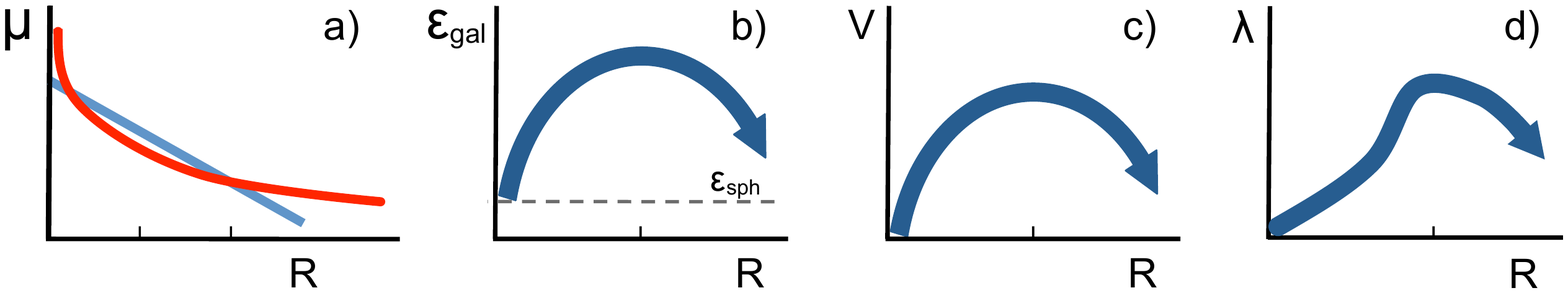}
\caption{ 
Panel a) shows the surface brightness profile of an ``ES'' galaxy with a S\'ersic
spheroid component (red) and an exponential intermediate-scale disk component
(blue).  Unlike large-scale disks in S0 galaxies, intermediate-scale disks do
not dominate the flux at large radii.  Panels b, c and d) roughly show how the
ellipticity, rotational velocity, and angular momentum profiles would behave.
}
\label{Fig8}
\end{figure*}

Measuring the rotation over roughly the inner 0.5--1.0 half-light galaxy
radii, 
$R_{\rm e,gal}$, of 260 ETGs, the ATLAS$^{3D}$ team realized that many galaxies
classified as elliptical had been misclassified because they rotated fast;
they contained disks (Emsellem et al.\ 2011; Krajnovi\'c et al.\ 2013).
Simply looking at images, and without a careful bulge/disk/etc.\ decomposition
of the galaxy light, it is easy to appreciate why many ES and S0 galaxies have
been misclassified as E galaxies.  The ATLAS$^{3D}$ team therefore continued
to use the earlier classification (Emsellem et al.\ 2007; Cappellari et
al.\ 2007) of ``fast rotator'' (FR) and ``slow rotator'' (SR) to replace the
photometrically determined S0 and E classification.  The limitation with this
new two-family dichotomy is that, while all ETG galaxies could previously be
accommodated for through the continuum from E to ES to S0, the ES galaxies
with intermediate-scale disks are not so adequately accounted for with the
FR/SR typing based on the kinematics within 1$R_{\rm e}$.  The issue is that
these somewhat overlooked galaxies are fast rotators at small to intermediate
radii and slow rotators at large radii.

\begin{figure}
\centering
\includegraphics[angle=0, trim=0.0cm 3.7cm 0.0cm 3.4cm, width=0.9\columnwidth]{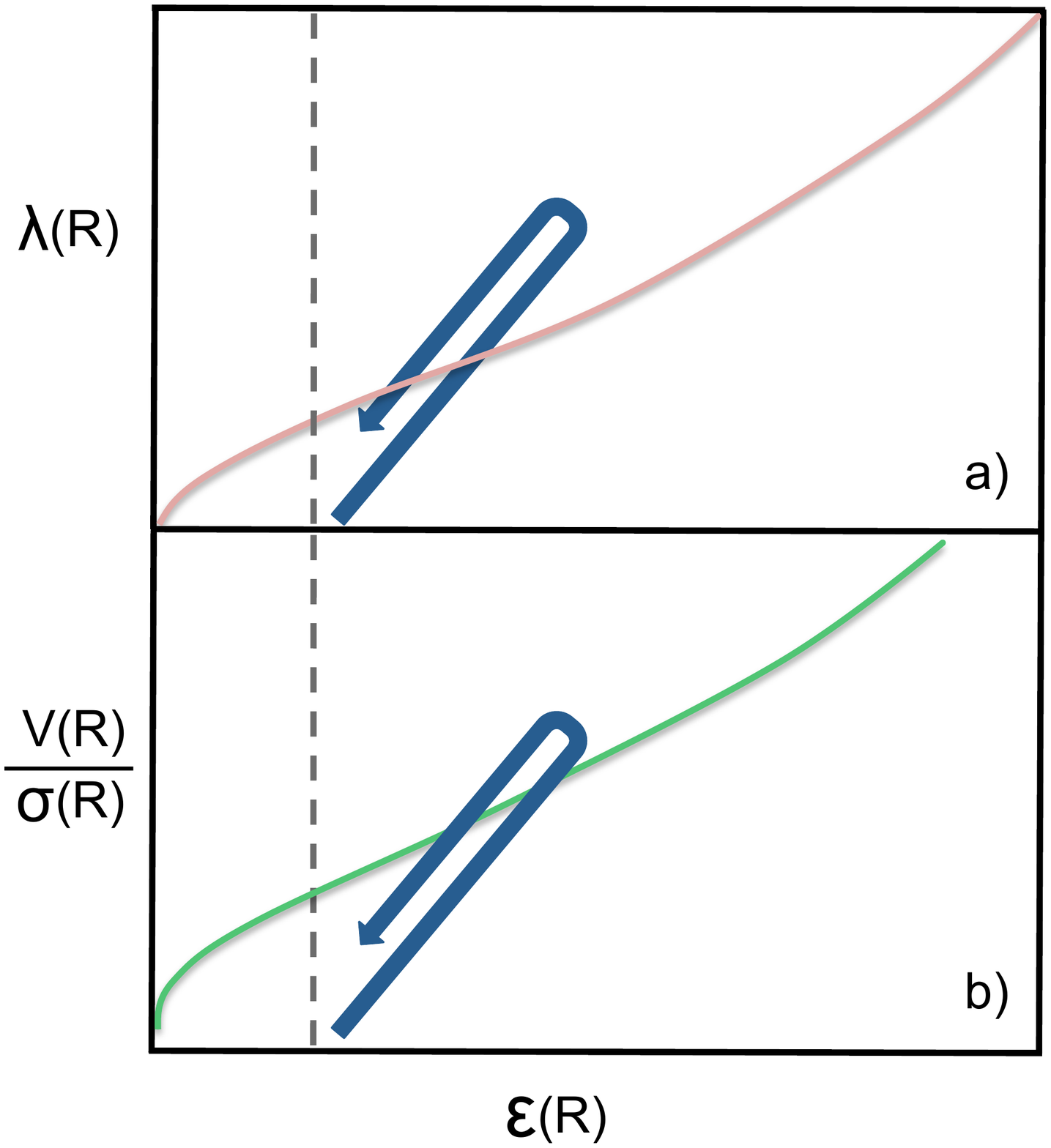} 
\caption{ Blue tracks show how a galaxy with an intermediate-scale disk
  might move as one samples the local angular momentum, $\lambda$, and the
  local rotational velocity (calibrated against the local or central velocity
  dispersion) as a function of increasing radius.  The dashed line denotes
  some fixed ellipticity for the spheroidal component of the galaxy, as shown
  in Figure~\ref{Fig8}.  Panel a): the pink curve approximates the edge-on view
  for ellipsoidal galaxies shown in Emsellem et al.\ (2007, their Fig.3),
  although this curve pertains to a luminosity-weighted spin parameter
  integrated within a radius of more than $\sim$$1R_{\rm e}$ and assumes an
  orbital anisotropy $\beta = 0.70 \times \epsilon$.  Panel b): the edge-on
  isotropic model from Cappellari et al.\ (2007, their Figure~3) is
  approximated here by the green curve.}
\label{Fig9}
\end{figure}

To parameterize, and thus better understand galaxies, 
Emsellem et al.\ (2011) replaced the luminosity-weighted $(V/\sigma)_{\rm
  e}$--$\epsilon_{\rm e}$ diagram (Emsellem et al.\ 2007) with the
luminosity-weighted $\lambda_{\rm e}$--$\epsilon_{\rm e}$ diagram. The
quantities involved, computed within the galaxy effective half light radius (hence the
subscripts `e'), are rotational velocity ($V$), velocity dispersion
($\sigma$), isophotal ellipticity ($\epsilon$), and a `spin' parameter defined
as  
\begin{equation}
\lambda_R = \frac{\sum_{i=1}^{N_p} F_i R_i \left| V_i
  \right|}{\sum_{i=1}^{N_p} F_i R_i \sqrt{V_i^2+\sigma_i^2}}, 
\end{equation}
with $F_i$ and $R_i$ the flux within, and geometric-mean radius 
of, the central then annular spatial bins out to some final radius $R$. 
Romanowsky \& Fall (2012) suggested replacing $\lambda_{\rm e} \equiv
\lambda_{R=R_{\rm e}}$ with the specific angular momenta, $j_* =J_*/M_*$,
where $J_*$ is the total angular momentum and $M_*$ is the total stellar mass
(Fall 1983; Casuso \& Beckman 2015, and references therein).  
While there are benefits to a more global parameter, instead of
a central ($R<1R_{\rm e}$) parameter, especially given that most of the
angular momenta can be at large radii, we propose and advocate capturing more
information than a single parameter like $\lambda_{\rm e}$ or $j_*$ or
$\epsilon$. 

Galaxies have been placed into a diagram of spin, $\lambda_{\rm e}$, versus a
single ellipticity parameter, $\epsilon$, and a dividing line used to type
them as FR or SR (e.g.\ Emsellem et al.\ 2011).  However, $\lambda$ is a
function of radius, and we suggest that this diagram not plot galaxies as
points, but instead plot them as tracks.  Sampling a galaxy at many radii can
result in a track across this diagram.  
Toloba et al.\ (2015) begin to demonstrate this in the dwarf regime by showing
results at 0.5$R_{\rm e}$ and 1$R_{\rm e}$.  
This is important because the ES
galaxies, but not the S0 galaxies,  
will transition from the FR region into the SR region of the diagram
as one starts to sample beyond the radial extent of their intermediate-scale
disk. 

Weighted by all the data within some radius $R$, the {\it cumulative} stellar
spin parameter $\lambda_R$ given above is less sensitive than the {\it local}
stellar spin parameter $\lambda(R)$ computed within an elliptical annulus of
radius $R$ (see Foster et al.\ 2016; Bellstedt et al.\ 2017a).  As such, radial profiles of
$\lambda(R)$ can be more revealing than radial profiles of $\lambda_R$.  In
addition, radial profiles of ellipticity are far more informative than a
single luminosity-weighted value within some radius.

Building on Liller (1966), 
Figure~\ref{Fig8} reveals what a galaxy with a somewhat edge-on
intermediate-scale disk may look like in radial plots of a) surface
brightness, b) ellipticity, c) velocity (or velocity divided by the central
velocity dispersion), and d) the local spin parameter.  Figure~\ref{Fig9} shows
how such a galaxy would move through the spin--ellipticity diagram as a
function of increasing galaxy radius.  

We advocate that galaxies should not be treated as single points in a stationary 
$\lambda_{R_{\rm e}}$--$\epsilon_{\rm e}$ diagram, 
but instead be treated as tracks moving across 
the $\lambda(R)$--$\epsilon(R)$ diagram. 
Given the shift to IFU surveys in recent years, this information can be readily
extracted.  Similarly, individual 
galaxy tracks in the $V(R)/\sigma(R)$--$\epsilon(R)$ diagram 
(e.g.\ Busarello et al.\ 1988, their Figure~11) would be 
more informative than points.   
While E galaxies should not move much, and S0 galaxies will quickly reach a
maximum, the ES galaxy tracks will be seen to recede from this maximum as one
probes out to larger radii. 
The use of `tracks' would therefore aid in the identification of the often 
overlooked, and subsequently mistreated, ES galaxies. 
It would, for example, enable a fuller understanding of the 
ETGs in the Calar Alto Legacy Integral Field Area (CALIFA) 
survey (S\'anchez et al.\ 2012).  From that team's 
multi-component decomposition of the light 
from 127 ETGs, M\'endez-Abreu et al.\ (2016) designated 36 of
them as ``B/BD'' systems, that is, they could not decide if these galaxies 
were either a pure bulge or a bulge$+$disk system.  
Many of these may be ellicular (ES) galaxies.  Excitingly, 
many of their galaxies have kinematic data out to 
several effective radii (e.g.\ Falc{\'o}n-Barroso et al.\ 2016), as does 
data from SAMI (Croom et al.\ 2012; Fogarty et al.\ 2015) 
and MaNGA (Bundy et al.\ 2015), enabling an 
exploration of fledgling disks and the transformation of elliptical galaxies
into lenticular galaxies.  

Unfortunately, we do not have enough radial extent in our kinematic data for
CG~611 to make an informative $\lambda_{R_{\rm e}}$--$\epsilon_{\rm e}$
diagram.  This may be the situation for many
dwarf galaxies due to their faint (and quickly declining) surface brightness
profiles.  However, we note that aside from dwarf ES (dES) galaxies, it will be
interesting to see the movement of ordinary (brighter) ES galaxies in the
$\lambda(R)$--$\epsilon(R)$ diagram.
Such ``disk ellipticals'' have been identified in Liller (1966), 
Nieto et al.\ (1988), Simien \& Michard (1990), Michard \& Marchal (1993), and
elsewhere.  Specific recent examples include 
NGC~3412 (Erwin et al.\ 2005, their Figure~3a), 
NGC~3115 (Arnold et 
al.\ 2011) and NGC~3377 (Arnold et al.\ 2014).  Other such galaxies of
interest that have intermediate-scale disks, include: NGC~1271 (Graham et al.\ 
2016); Mrk~1216 and NGC~1332 (Savorgnan \& Graham 2016b);
NGC~821, NGC~3377, and NGC~4697 (Savorgnan \& Graham 2016a); and NGC~4473
(Foster et al.\ 2013; Alabi et al.\ 2015).   
Galaxies like NGC~1277 (Graham et al.\ 2016) and NGC~3115 (Savorgnan \& Graham
2016b) also have disks fully embedded in their spheroids, but they extend
quite far out and thus globular clusters or planetary nebula, rather than
long-slit or integral field unit (IFU) spectroscopy will likely be required to
sample the dynamics beyond the disk (e.g.\ 
Hui et
al.\ 1995; Romanowsky et al.\ 2003; Teodorescu et al.\ 2005; Woodley et
al.\ 2007; Pota et al.\ 2013; Brodie et al.\ 2014; Richtler et al.\ 2015;
Toloba et al.\ 2016), though this will be challenging for CG~611 at a
`luminosity distance' of 46.7 Mpc.

\subsubsection{$S^2_{K}$ and the Fundamental Plane}\label{Sec_Sk} 

The existence of galaxies with varyingly sized disks raises a second issue 
that we address here. 

To account for the contribution to kinetic energy coming from not just the
random motion of stars, 
as traced by the velocity dispersion $\sigma$, but
also from the ordered motion given by the rotational velocity $V_{\rm rot}$, 
Busarello et al.\ (1992) effectively used the quantity 
\begin{equation} 
S^2_{1/3} \equiv \frac{3}{2}\sigma^2 + \frac{1}{2}V_{\rm rot}^2, \nonumber
\end{equation}
where 
$3\sigma^2$ represents the sum of the velocity dispersion squared in each plane 
\begin{equation} 
\sigma^2_{XY} + \sigma^2_{XZ} + \sigma^2_{YZ}. 
\end{equation} 
The existence of orbital anisotropy means that these last three quantities may
not be equal. 
The quantity $S^2_{1/3}$ reduces to 1.5$\sigma^2$ if there is no rotation in a
galaxy, and reduces to 0.5$V_{\rm rot}^2$ if there is only a rotating disk
with no velocity dispersion.  The important 1 to 3 scaling reflects that
rotation predominantly occurs in one plane while velocity dispersion occurs in all three
planes. 

The above quantity is generalized in Weiner et al.\ (2006), such that 
\begin{equation}
S^2_{K} \equiv \sigma^2 + KV_{\rm rot}^2, \nonumber
\end{equation} 
which is equivalent to the first expression when $K=1/3$.  As can be seen, the
quantity $K$ sets the ratio between the ordered and random motions in this
generalized kinematic scaling expression $S^2_K$.

It would appear logical that studies building on the ``Fundamental Plane''
(FP: Djorgovski \& Davis 1987), which are now using $S^2_{1/2}$ rather than just
$\sigma^2$ for the ETGs (and for spiral galaxies, e.g.\ Kassin et al.\ 2007;
Zaritsky et al.\ 2008; Cortese et al.\ 2014), may benefit from a recognition
of ES galaxies in their surveys and a modified treatment of them.

The virial theorem, considered to be the basis for the FP, relates the mass
$M$ of a system to the speeds and sizes of its components.  At a galaxy
component level, 
\begin{equation}
M \approx \frac{3}{2}\sigma^2R_{\rm sph} + \frac{1}{2}V_{\rm rot}^2R_{\rm
  disk}. 
\end{equation} 
For disk galaxies, the ratio of the spheroid's effective half-light radius to
the disk's scalelength is known to be fairly constant, with 
$R_{\rm e}/h \approx 0.2$ (e.g.\ Courteau et al.\ 1996; Graham \& 
Worley 2008, and references therein).  However, in ES galaxies, the size of
the disk relative to the spheroid is not constant.  Although many FP-type
studies are now accounting for the different kinematics of the two major
components in galaxies, i.e.\ spheroid and disk, they do not yet account for the
different physical sizes of these components.  This may be an issue for
studies of ETGs including both S0 and ES galaxies. 
Rather than simply using, or implicitly assuming, 
a single galaxy size --- typically the `effective half-light 
radius' $R_{\rm e,gal}$ --- the use of appropriate galaxy component sizes may help
to bring increased physical meaning to the photometric and spectroscopic survey
data.  Doing so may lead to a deeper physical understanding of galaxies, 
possibly reduce some of the scatter about the modified Fundamental Plane, and
in turn lead to an improved distance indicator for studies of peculiar velocity
flows.

\section{Conclusions} 

CG~611 is a dwarf early-type galaxy that formed in isolation (nature,
rather than nurture by a parent galaxy cluster or a neighboring `big brother' 
galaxy).  It has experienced gas accretion which contributed to 
the build-up of the inner disk.  This ETG's ionized gas is
observed to counter-rotate with respect to the majority of stars in the inner
$\approx$1 $R_{\rm e,gal}$.  Such a scenario is mentioned in Koleva et
al.\ (2013), and it has resulted in the creation of this dwarf 
galaxy.  CG~611's isolation, and accreted gas disk, 
reveals that it is not a late-type galaxy morphologically
transformed by a cluster environment, i.e.\ the argument advocated for years
for the creation of the rotating dwarf ETGs.  If CG~611
were to fall into a cluster, ram pressure stripping would remove its gas and
dust, and it would then immediately resemble the dwarf ETGs already in clusters.

Having taken the `cluster environment' out of the equation, CG~611 
reveals that it is accretion, the build-up of gas from the primordial supply, 
that can give dwarf ETGs their spin 
(e.g.\ Jin et al.\ 2016).  That is, rotating dwarf ETGs are not
necessarily built from the removal of material (from late-type galaxies) in a
cluster --- although some likely are. 

To better understand the nature of CG~611, it would be helpful to: 
$i)$ obtain an {\it HST} image, enabling
one to quantify additional components near the center of this galaxy and
better constrain the suspected intermediate-scale disk, bar and lens; 
$ii)$ see what, if any, hydrogen or molecular gas cloud is 
associated with CG~611, as this should provide additional
information about the disk formation in this galaxy; 
$iii)$ acquire kinematical information at larger radii, perhaps by probing the
globular cluster system, to determine the ratio of ordered rotation to random
motion of the main galaxy beyond the inner $R_{\rm e,gal}$; 
and 
$iv)$ check for any AGN X-ray emission coming from this galaxy, because it is a good
candidate to host an intermediate mass black hole.

In several aspects, CG~611 resembles a number of other dwarf ETGs, in
particular, VCC~216, VCC~490, and NGC~4150.  
In regards to two other possibly similar galaxies, 
it would be interesting to check for the existence of an inner spiral in the
isolated compact elliptical galaxy J094729.24+141245.3 (Huxor et al.\ 2013).
This galaxy may be an additional example of ETG evolution via disk growth
rather than disk stripping.
It will also be interesting to obtain spatially resolved kinematics for
CGCG~036-042 (Paudel et al.\ 2014) with its 4 Gyr old inner exponential
component, and thereby answer what rotation, if any, is associated with this
component. 

The disk in CG~611, which is undergoing development, has motivated us to 
advocate representing galaxies by tracks, 
rather than single points, in a revised spin--ellipticity diagram. 
Although we do not yet have sufficiently radially extended kinematics to do this
for CG~611, Bellstedt et al. (2017b) presents such tracks for a
number of brighter ``disk elliptical'' ES galaxies. 
We have additionally advocated the consideration of disk and spheroid sizes when using
the disk and spheroid kinematics for the quantity $S_K = \sqrt{K\,V^2_{\rm
    rot} + \sigma^2}$, which may prove useful for ``Fundamental Plane'' type studies.

\acknowledgments

This research was supported under the Australian Research Council's funding
scheme (DP17012923), and through Swinburne's Keck time Project Codes W028E and W138E. 
This research has made use of the NASA/IPAC Extragalactic Database (NED), 
the SAO/NASA Astrophysics Data System, and the SDSS archives.


\begin{thebibliography}{}
\bibitem[Abazajian et al.(2005)]{2005AJ....129.1755A} Abazajian, K.,
  Adelman-McCarthy, J.~K., Ag{\"u}eros, M.~A., et al.\ 2005, \aj, 129, 1755
\bibitem[Aguerri(2016)]{2016A&A...587A.111A} Aguerri, J.~A.~L.\ 2016, \aap,
  587, A111 
\bibitem[Alabi et al.(2015)]{2015MNRAS.452.2208A} Alabi, A.~B., Foster, C.,
  Forbes, D.~A., et al.\ 2015, \mnras, 452, 2208 
\bibitem[Alatalo et al.(2013)]{2013MNRAS.432.1796A} Alatalo, K., Davis, T.~A.,
  Bureau, M., et al.\ 2013, \mnras, 432, 1796 
\bibitem[Ann et al.(2015)]{2015ApJS..217...27A} Ann, H.~B., Seo, M., \& Ha,
  D.~K.\ 2015, \apjs, 217, 27 
\bibitem[Argudo-Fern{\'a}ndez et al.(2015)]{2015A&A...578A.110A}
  Argudo-Fern{\'a}ndez, M., Verley, S., Bergond, G., et al.\ 2015, \aap, 578,
  A110 
\bibitem[Arnold et al.(2011)]{2011ApJ...736L..26A} Arnold, J.~A., Romanowsky,
  A.~J., Brodie, J.~P., et al.\ 2011, \apjl, 736, L26 
\bibitem[Arnold et al.(2014)]{2014ApJ...791...80A} Arnold, J.~A., Romanowsky,
  A.~J., Brodie, J.~P., et al.\ 2014, \apj, 791, 80 
\bibitem[Barazza et al.(2002)]{2002A&A...391..823B} Barazza, F.~D., Binggeli,
  B., \& Jerjen, H.\ 2002, \aap, 391, 823 
\bibitem[Begelman et al.(1980)]{1980Natur.287..307B} Begelman, M.~C.,
   Blandford, R.~D., \& Rees, M.~J.\ 1980, \nat, 287, 307 
\bibitem[Bekki et al.(2001)]{2001ApJ...557L..39B} Bekki, K., Couch, W.~J.,
   Drinkwater, M.~J., \& Gregg, M.~D.\ 2001, \apjl, 557, L39 
\bibitem[Bellstedt et al.(2017a)]{Bellstedt}Bellstedt, S., Forbes, D.A.,
  Foster, C., et al.\ 2017a, MNRAS, 467, 4540
\bibitem[Bellstedt et al.(2017b)]{Bellstedt}Bellstedt, S., Graham, A.W.,
  Forbes, D.A., et al.\ 2017b, submitted
\bibitem[Bender et al.(1992)]{1992ApJ...399..462B} Bender, R., Burstein, D.,
  \& Faber, S.~M.\ 1992, \apj, 399, 462 
\bibitem[Benson et al.(2015)]{2015ApJ...799..171B} Benson, A.~J., Toloba, E.,
  Mayer, L., Simon, J.~D., \& Guhathakurta, P.\ 2015, \apj, 799, 171 
\bibitem[Bettoni et al.(1988)]{1988A&A...197...69B} Bettoni, D., Galletta,
   G., \& Vallenari, A.\ 1988, \aap, 197, 69 
\bibitem[Bialas et al.(2015)]{2015A&A...576A.103B} Bialas, D., Lisker, T.,
  Olczak, C., Spurzem, R., \& Kotulla, R.\ 2015, \aap, 576, A103
\bibitem[Birnboim \& Dekel(2003)]{2003MNRAS.345..349B} Birnboim, Y., \&
   Dekel, A.\ 2003, MNRAS, 345, 349
\bibitem[Blanton et al.(2011)]{2011AJ....142...31B} Blanton, M.~R., Kazin, E.,
  Muna, D., Weaver, B.~A., \& Price-Whelan, A.\ 2011, \aj, 142, 31
\bibitem[Boselli et al.(2008)]{2008ApJ...674..742B} Boselli, A., Boissier, S.,
  Cortese, L., \& Gavazzi, G.\ 2008, \apj, 674, 742
\bibitem[Boselli et al.(2016)]{2016A&A...587A..68B} Boselli, A., Cuillandre,
  J.~C., Fossati, M., et al.\ 2016, \aap, 587, A68 
\bibitem[Brodie et al.(2014)]{2014ApJ...796...52B} Brodie, J.~P., Romanowsky,
  A.~J., Strader, J., et al.\ 2014, \apj, 796, 52 
\bibitem[Brooks et al.(2009)]{2009ApJ...694..396B} Brooks, A.~M., Governato,
  F., Quinn, T., Brook, C.~B., \& Wadsley, J.\ 2009, \apj, 694, 396 
\bibitem[Bryant et al.(2015)]{2015MNRAS.447.2857B} Bryant, J.~J., Owers,
  M.~S., Robotham, A.~S.~G., et al.\ 2015, \mnras, 447, 2857 
\bibitem[Bundy et al.(2015)]{2015ApJ...798....7B} Bundy, K., Bershady, M.~A.,
  Law, D.~R., et al.\ 2015, \apj, 798, 7 
\bibitem[Busarello et al.(1988)]{1988A&A...197...91B} Busarello, G., Filippi,
   S., \& Ruffini, R.\ 1988, \aap, 197, 91 
\bibitem[Busarello et al.(1992)]{1992A&A...262...52B} Busarello, G., Longo,
   G., \& Feoli, A.\ 1992, \aap, 262, 52 
\bibitem[Buta \& Crocker(1991)]{1991AJ....102.1715B} Buta, R., \& Crocker,
   D.~A.\ 1991, \aj, 102, 1715 
\bibitem[Butcher \& Oemler(1978)]{1978ApJ...226..559B} Butcher, H., \& Oemler,
  A., Jr.\ 1978, \apj, 226, 559 
\bibitem[Cair{\'o}s et al.(2001a)]{2001ApJS..133..321C} Cair{\'o}s, L.~M.,
  V{\'{\i}}lchez, J.~M., Gonz{\'a}lez P{\'e}rez, J.~N., Iglesias-P{\'a}ramo,
  J., \& Caon, N.\ 2001a, \apjs, 133, 321 
\bibitem[Cair{\'o}s et al.(2001b)]{2001ApJS..136..393C} Cair{\'o}s, L.~M.,
  Caon, N., V{\'{\i}}lchez, J.~M., Gonz{\'a}lez-P{\'e}rez, J.~N., \&
  Mu{\~n}oz-Tu{\~n}{\'o}n, C.\ 2001b, \apjs, 136, 393 
\bibitem[Cair{\'o}s et al.(2002)]{2002ApJ...577..164C} Cair{\'o}s, L.~M.,
  Caon, N., Garc{\'{\i}}a-Lorenzo, B., V{\'{\i}}lchez, J.~M., \&
  Mu{\~n}oz-Tu{\~n}{\'o}n, C.\ 2002, \apj, 577, 164 
\bibitem[Caon et al.(2005)]{2005ApJS..157..218C} Caon, N., Cair{\'o}s, L.~M.,
  Aguerri, J.~A.~L., \& Mu{\~n}oz-Tu{\~n}{\'o}n, C.\ 2005, \apjs, 157, 218 
\bibitem[Cappellari \& Emsellem(2004)]{2004PASP..116..138C} Cappellari, M., \&
  Emsellem, E.\ 2004, \pasp, 116, 138 
\bibitem[Cappellari et al.(2007)]{2007MNRAS.379..418C} Cappellari, M.,
  Emsellem, E., Bacon, R., et al.\ 2007, \mnras, 379, 418 
\bibitem[Casuso \& Beckman(2015)]{2015MNRAS.449.2910C} Casuso, E., \&
   Beckman, J.~E.\ 2015, \mnras, 449, 2910 
\bibitem[Chilingarian et al.(2009)]{2009Sci...326.1379C} Chilingarian, I.,
   Cayatte, V., Revaz, Y., et al.\ 2009, Science, 326, 1379 
\bibitem[Chilingarian \& Zolotukhin(2015)]{2015Sci...348..418C} Chilingarian,
  I., \& Zolotukhin, I.\ 2015, Science, 348, 418 
\bibitem[Chilingarian et al.(2017)]{Chili:2016} Chilingarian, I.~V.,
  Zolotukhin, I., Katkov, I., Melchior, A.-L.\ 2017, ApJS, 228, 14 
\bibitem[Christensen et al.(2016)]{2016ApJ...824...57C} Christensen, C.~R.,
  Dav{\'e}, R., Governato, F., et al.\ 2016, \apj, 824, 57
\bibitem[Ciambur(2015)]{2015ApJ...810..120C} Ciambur, B.~C.\ 2015, \apj, 810,
   120 
\bibitem[Ciambur(2017)]{2016arXiv160708620C} Ciambur, B.~C.\ 2017, PASA, 33, 62
\bibitem[Collobert et al.(2006)]{2006MNRAS.370.1213C} Collobert, M., Sarzi,
   M., Davies, R.~L., Kuntschner, H., \& Colless, M.\ 2006, \mnras, 370, 1213 
\bibitem[Cornuault et al.(2016)]{Cornuault:2016} Cornuault, N., Lehnert, M.D.,
  Boulanger, F., Guillard, P.\ 2016, A\&A, submitted (arXiv:1609.04405) 
\bibitem[Corsini(2014)]{2014ASPC..486...51C} Corsini, E.~M.\ 2014, Multi-Spin
   Galaxies, ASP Conference Series, 486, 51 
\bibitem[Cortese et al.(2014)]{2014ApJ...795L..37C} Cortese, L., Fogarty,
   L.~M.~R., Ho, I.-T., et al.\ 2014, \apjl, 795, L37 
\bibitem[C{\^o}t{\'e} et al.(2007)]{2007ApJ...671.1456C} C{\^o}t{\'e}, P.,
   Ferrarese, L., Jord{\'a}n, A., et al.\ 2007, \apj, 671, 1456 
\bibitem[C{\^o}t{\'e} et al.(2008)]{2008IAUS..245..395C} Côté, P., Ferrarese,
  L., Jordán, A., et al.\ 2008, in IAU Symp.\ 245, Formation \& Evolution of
  Galaxy Bulges, ed.\ M.\ Bureau, E.\ Athanassoula \& B.\ Barbuy (Cambridge:
  Cambridge Univ.\ Press), 395 
\bibitem[Courteau et al.(1996)]{1996ApJ...457L..73C} Courteau, S., de Jong,
   R.~S., \& Broeils, A.~H.\ 1996, \apjl, 457, L73 
\bibitem[Croom et al.(2012)]{2012MNRAS.421..872C} Croom, S.~M., Lawrence,
  J.~S., Bland-Hawthorn, J., et al.\ 2012, \mnras, 421, 872 
\bibitem[Danovich et al.(2012)]{2012MNRAS.422.1732D} Danovich, M., Dekel, A.,
   Hahn, O., \& Teyssier, R.\ 2012, MNRAS, 422, 1732
\bibitem[Danovich et al.(2014)]{2014arXiv1407.7129D} Danovich, M., Dekel, A.,
   Hahn, O., Ceverino, D., \& Primack, J.\ 2014, MNRAS, 449, 2087
\bibitem[Davies et al.(1983)]{1983ApJ...266...41D} Davies, R.~L., Efstathiou,
  G., Fall, S.~M., Illingworth, G., \& Schechter, P.~L.\ 1983, \apj, 266, 41 
\bibitem[Davis et al.(2011)]{2011MNRAS.417..882D} Davis, T.~A., Alatalo, K.,
  Sarzi, M., et al.\ 2011, \mnras, 417, 882 
\bibitem[Dekel \& Birnboim(2006)]{2006MNRAS.368....2D} Dekel, A., \& Birnboim,
  Y.\ 2006, \mnras, 368, 2 
\bibitem[Dekel et al.(2009)]{Dek09}Dekel, A., Birnboim, Y., Engel, G., et
  al.\ 2009, Nature, 457, 451
\bibitem[de la Rosa et al.(2016)]{2016MNRAS.457.1916D} de la Rosa, I.~G., La
  Barbera, F., Ferreras, I., et al.\ 2016, \mnras, 457, 1916 
\bibitem[De Lucia et al.(2009)]{2009MNRAS.400...68D} De Lucia, G., Poggianti,
   B.~M., Halliday, C., et al.\ 2009, \mnras, 400, 68 
\bibitem[den Brok et al.(2011)]{2011MNRAS.414.3052D} den Brok, M., Peletier,
  R.~F., Valentijn, E.~A., et al.\ 2011, \mnras, 414, 3052 
\bibitem[de Vaucouleurs(1957)]{deV57}de Vaucouleurs, G.\ 1957, AJ, 62, 69
\bibitem[D'Onofrio et al.(2015)]{2015FrASS...2....4D} D'Onofrio, M., Marziani,
  P., \& Buson, L.\ 2015, Frontiers in Astronomy and Space Sciences, 2, 4 
\bibitem[de Rijcke et al.(2005)]{2005A&A...438..491D} de Rijcke, S.,
   Michielsen, D., Dejonghe, H., Zeilinger, W.~W., \& Hau, G.~K.~T.\ 2005,
  \aap, 438, 491 
\bibitem[De Rijcke et al.(2010)]{2010ApJ...724L.171D} De Rijcke, S., Van Hese,
  E., \& Buyle, P.\ 2010, \apjl, 724, L171 
\bibitem[Dullo \& Graham(2014)]{2014MNRAS.444.2700D} Dullo, B.~T., \& Graham,
   A.~W.\ 2014, \mnras, 444, 2700 
\bibitem[Djorgovski \& Davis(1987)]{1987ApJ...313...59D} Djorgovski, S., \&
   Davis, M.\ 1987, \apj, 313, 59 
\bibitem[Ebeling et al.(2014)]{2014ApJ...781L..40E} Ebeling, H., Stephenson,
  L.~N., \& Edge, A.~C.\ 2014, \apjl, 781, L40 
\bibitem[Elmegreen \& Struck(2016)]{Elmegreen}Elmegreen, B.G., Struck,
  C.\ 2016, ApJ, 830, 115
\bibitem[Emsellem et al.(2007)]{2007MNRAS.379..401E} Emsellem, E., Cappellari,
  M., Krajnovi{\'c}, D., et al.\ 2007, \mnras, 379, 401 
\bibitem[Emsellem et al.(2011)]{2011MNRAS.414..888E} Emsellem, E., Cappellari,
  M., Krajnovi{\'c}, D., et al.\ 2011, \mnras, 414, 888 
\bibitem[Erwin et al.(2005)]{2005ApJ...626L..81E} Erwin, P., Beckman, J.~E.,
   \& Pohlen, M.\ 2005, \apjl, 626, L81 
\bibitem[Faber(1973)]{1973ApJ...179..423F} Faber, S.~M.\ 1973, \apj, 179, 423 
\bibitem[Falc{\'o}n-Barroso et al.(2016)]{2016arXiv160906446F}
   Falc{\'o}n-Barroso, J., Lyubenova, M., van de Ven, G., et al.\ 2016, A\&A,
   597, A48
\bibitem[Fall(1983)]{1983IAUS..100..391F} Fall, S.~M.\ 1983, in IAU Symp.\ 100,
  Internal Kinematics and Dynamics of Galaxies, ed.\ E.\ Athanassoula
  (Cambridge: Cambridge Univ.\ Press), 391 
\bibitem[Ferrarese et al.(2006)]{2006ApJS..164..334F} Ferrarese, L.,
   C{\^o}t{\'e}, P., Jord{\'a}n, A., et al.\ 2006, \apjs, 164, 334 
\bibitem[Ferrarese(2016)]{2016:Ferrarese}Ferrarese, L.\ 2016, in ``From the
   Realm of the Nebulae to Populations of Galaxies: Dialogues on a century of
   Research'', M.\ D'Onofrio, R.\ Rampazzo, S.\ Zaggia (Eds.), Springer
   Publishing, Astrophysics and Space Science Library, v.435, p.402-407
\bibitem[Ferrers(1877)]{F1877}Ferrers N.M.\ 1877, Quart.\ J.\ Pure Appl.\ Math., 14, 1
\bibitem[Fogarty et al.(2015)]{2015MNRAS.454.2050F} Fogarty, L.~M.~R., Scott,
  N., Owers, M.~S., et al.\ 2015, \mnras, 454, 2050 
\bibitem[Forbes et al.(1996)]{1996ApJ...467..126F} Forbes, D.~A., Franx, M.,
   Illingworth, G.~D., \& Carollo, C.~M.\ 1996, \apj, 467, 126 
\bibitem[Forbes et al.(2008)]{2008MNRAS.389.1924F} Forbes, D.~A., Lasky, P.,
  Graham, A.~W., \& Spitler, L.\ 2008, \mnras, 389, 1924 
\bibitem[Foster et al.(2013)]{2013MNRAS.435.3587F} Foster, C., Arnold, J.~A.,
  Forbes, D.~A., et al.\ 2013, \mnras, 435, 3587 
\bibitem[Foster et al.(2016)]{2016MNRAS.457..147F} Foster, C., Pastorello,
   N., Roediger, J., et al.\ 2016, \mnras, 457, 147 
\bibitem[Freeman(1970)]{Fre70}Freeman, K.C.\ 1970, ApJ, 160, 811
\bibitem[Fukugita et al.(1995)]{1995PASP..107..945F} Fukugita, M., Shimasaku,
  K., \& Ichikawa, T.\ 1995, \pasp, 107, 945 
\bibitem[Fuse et al.(2012)]{2012AJ....144...57F} Fuse, C., Marcum, P., \&
  Fanelli, M.\ 2012, \aj, 144, 57 
\bibitem[Galletta(1987)]{1987ApJ...318..531G} Galletta, G.\ 1987, \apj, 318,
   531 
\bibitem[Gavazzi et al.(2001)]{2001ApJ...563L..23G} Gavazzi, G., Boselli, A.,
  Mayer, L., et al.\ 2001, \apjl, 563, L23 
\bibitem[Gavazzi et al.(2004)]{2004A&A...417..499G} Gavazzi, G., Zaccardo, A.,
  Sanvito, G., Boselli, A., \& Bonfanti, C.\ 2004, \aap, 417, 499 
\bibitem[Geha et al.(2003)]{2003AJ....126.1794G} Geha, M., Guhathakurta, P.,
  \& van der Marel, R.~P.\ 2003, \aj, 126, 1794 
\bibitem[Geha et al.(2010)]{2010ApJ...711..361G} Geha, M., van der Marel,
  R.~P., Guhathakurta, P., et al.\ 2010, \apj, 711, 361 
\bibitem[Genel et al.(2010)]{2010ApJ...719..229G} Genel, S., Bouch{\'e}, N.,
  Naab, T., Sternberg, A., \& Genzel, R.\ 2010, \apj, 719, 229 
\bibitem[Gill et al.(2005)]{2005MNRAS.356.1327G} Gill, S.~P.~D., Knebe, A.,
   \& Gibson, B.~K.\ 2005, \mnras, 356, 1327 
\bibitem[Gonz{\'a}lez-Samaniego et al.(2014)]{2014ApJ...785...58G}
  Gonz{\'a}lez-Samaniego, A., Col{\'{\i}}n, P., Avila-Reese, V.,
  Rodr{\'{\i}}guez-Puebla, A., \& Valenzuela, O.\ 2014, \apj, 785, 58 
\bibitem[Graham(2002)]{2002ApJ...568L..13G} Graham, A.~W.\ 2002, \apjl, 568,
   L13 
\bibitem[Graham(2005)]{2005nfcd.conf..303G}Graham, A.W.\ 2005, in IAU
  Colloq.\ 198, Near-fields Cosmology with Dwarf Elliptical Galaxies,
  ed.\ H.\ Jerjen \& B.\ Binggeli (Cambridge: Cambridge Univ.\ Press), 303
\bibitem[Graham(2013)]{Graham:2013}Graham, A.W.\ 2013, in ``Planets, Stars
   and Stellar Systems'', Volume 6, p.91-140, T.D.Oswalt \& W.C.Keel (Eds.),
   Springer Publishing (arXiv:1108.0997)
\bibitem[Graham(2016)]{Graham:2016}Graham, A.W.\ 2016, in ``Galactic
   Bulges'', E.\ Laurikainen, R.F.\ Peletier \& D.\ Gadotti (Eds.), Springer
   International Publishing, Astrophysics and Space Science Library, v.418,
   p.263-313 (arXiv:1501.02937) 
\bibitem[Graham et al.(2016)]{2016arXiv160800711G} Graham, A.~W., Ciambur,
  B.~C., \& Savorgnan, G.~A.~D.\ 2016, ApJ, 831, 132 
\bibitem[Graham et al.(1998)]{1998A&AS..133..325G} Graham, A.~W., Colless,
  M.~M., Busarello, G., Zaggia, S., \& Longo, G.\ 1998, \aaps, 133, 325 
\bibitem[Graham et al.(2015)]{2015ApJ...804...32G} Graham, A.~W., Dullo,
  B.~T., \& Savorgnan, G.~A.~D.\ 2015, \apj, 804, 32 
\bibitem[Graham \& Guzm{\'a}n(2003)]{2003AJ....125.2936G} Graham, A.~W., \&
  Guzm{\'a}n, R.\ 2003, \aj, 125, 2936 
\bibitem[Graham \& Guzm\'an(2004)]{2004ASSL..319..723G} Graham, A.~W., \&
  Guzm\'an, R.\ 2004, in Penetrating Bars Through Masks of Cosmic Dust, Edited
  by D.L.\ Block, I.\ Puerari, K.C.\ Freeman, R.\ Groess, and E.K.\ Block.
  Dordrecht: Kluwer Academic Publishers, ASSL, 319, 723 
\bibitem[Graham et al.(2003)]{2003AJ....126.1787G} Graham, A.~W., Jerjen, H.,
  \& Guzm{\'a}n, R.\ 2003, \aj, 126, 1787 
\bibitem[Graham et al.(2006)]{2006AJ....132.2711G} Graham, A.~W., Merritt, D.,
  Moore, B., Diemand, J., \& Terzi{\'c}, B.\ 2006, \aj, 132, 2711 
\bibitem[Graham et al.(2012)]{2012ApJ...750..121G} Graham, A.~W., Spitler,
  L.~R., Forbes, D.~A., et al.\ 2012, \apj, 750, 121 
\bibitem[Graham \& Worley(2008)]{2008MNRAS.388.1708G} Graham, A.~W., \&
   Worley, C.~C.\ 2008, \mnras, 388, 1708 
\bibitem[Grebel et al.(2003)]{2003AJ....125.1926G} Grebel, E.~K., Gallagher,
   J.~S., III, \& Harbeck, D.\ 2003, \aj, 125, 1926 
\bibitem[Gunn \& Gott(1972)]{1972ApJ...176....1G} Gunn, J.~E., \& Gott, J.~R.,
  III 1972, \apj, 176, 1 
\bibitem[Held et al.(1992)]{1992AJ....103..851H} Held, E.~V., de Zeeuw, T.,
  Mould, J., \& Picard, A.\ 1992, \aj, 103, 851 
\bibitem[Hern{\'a}ndez-Toledo et al.(2010)]{2010AJ....139.2525H}
  Hern\'andez-Toledo, H.~M., V\'azquez-Mata, J.~A.,
  Mart{\'{\i}}nez-V\'azquez, L.~A., Choi, Y.-Y., \& Park, C.\ 2010, \aj,
  139, 2525 
\bibitem[Ho et al.(2011)]{2011ApJS..197...21H} Ho, L.~C., Li, Z.-Y., Barth,
   A.~J., Seigar, M.~S., \& Peng, C.~Y.\ 2011, \apjs, 197, 21 
\bibitem[Huchra et al.(2012)]{2012ApJS..199...26H} Huchra, J.~P., Macri,
   L.~M., Masters, K.~L., et al.\ 2012, \apjs, 199, 26 
\bibitem[Hui et al.(1995)]{1995ApJ...449..592H} Hui, X., Ford, H.~C.,
   Freeman, K.~C., \& Dopita, M.~A.\ 1995, \apj, 449, 592 
\bibitem[Huxor et al.(2013)]{2013MNRAS.430.1956H} Huxor, A.~P., Phillipps, S.,
  \& Price, J.\ 2013, \mnras, 430, 1956 
\bibitem[Janz \& Lisker(2008)]{2008ApJ...689L..25J} Janz, J., \& Lisker,
  T.\ 2008, \apjl, 689, L25 
\bibitem[Janz \& Lisker(2009)]{2009ApJ...696L.102J} Janz, J., \& Lisker,
  T.\ 2009, \apjl, 696, L102 
\bibitem[Janz et al.(2012)]{2012ApJ...745L..24J} Janz, J., Laurikainen, E.,
  Lisker, T., et al.\ 2012, \apjl, 745, L24 
\bibitem[Janz et al.(2016)]{2016MNRAS.461L..82J} Janz, J., Laurikainen, E.,
  Laine, J., Salo, H., \& Lisker, T.\ 2016, \mnras, 461, L82 
\bibitem[Janz et al.(2017)]{Janz:2017}Janz, J., Penny, S.J., Graham, A.W.,
  Forbes, D.A., Davies, R.L.\ 2017, MNRAS, in press (arXiv:1703.04975)
\bibitem[Jarrett et al.(2000)]{2MASS}Jarrett, T.H., Chester, T., Cutri, R., et
  al.\ 2000, AJ, 119, 2498 (2MASS)
\bibitem[Jarrett et al.(2011)]{2011ApJ...735..112J} Jarrett, T.~H., Cohen,
   M., Masci, F., et al.\ 2011, \apj, 735, 112 
\bibitem[Jerjen et al.(2000)]{2000A&A...358..845J} Jerjen, H., Kalnajs, A., \&
  Binggeli, B.\ 2000, \aap, 358, 845 
\bibitem[Jin et al.(2016)]{2016MNRAS.463..913J} Jin, Y., Chen, Y., Shi, Y., et
  al.\ 2016, \mnras, 463, 913 
\bibitem[Karachentseva et al.(2010)]{2010Ap.....53..462K} Karachentseva,
  V.~E., Karachentsev, I.~D., \& Sharina, M.~E.\ 2010, Astrophysics, 53, 462
\bibitem[Kassin et al.(2007)]{2007ApJ...660L..35K} Kassin, S.~A., Weiner,
   B.~J., Faber, S.~M., et al.\ 2007, \apjl, 660, L35 
\bibitem[Katkov et al.(2014)]{2014MNRAS.438.2798K} Katkov, I.~Y., Sil'chenko,
  O.~K., \& Afanasiev, V.~L.\ 2014, \mnras, 438, 2798
\bibitem[Katkov et al.(2015)]{2015AJ....150...24K} Katkov, I.~Y., Kniazev,
  A.~Y., \& Sil'chenko, O.~K.\ 2015, \aj, 150, 24 
\bibitem[Kazantzidis et al.(2011)]{2011ApJ...726...98K} Kazantzidis, S.,
  {\L}okas, E.~L., Callegari, S., Mayer, L., \& Moustakas, L.~A.\ 2011, \apj,
  726, 98
\bibitem[Kazantzidis et al.(2013)]{2013ApJ...764L..29K} Kazantzidis, S.,
  {\L}okas, E.~L., \& Mayer, L.\ 2013, \apjl, 764, L29 
\bibitem[Keenan \& Innanen(1975)]{1975AJ.....80..290K} Keenan, D.~W., \&
   Innanen, K.~A.\ 1975, \aj, 80, 290 
\bibitem[Kere{\v s} et al.(2005)]{2005MNRAS.363....2K} Kere{\v s}, D., Katz,
  N., Weinberg, D.~H., \& Dav{\'e}, R.\ 2005, MNRAS, 363, 2
\bibitem[King(1962)]{1962AJ.....67..471K} King, I.\ 1962, \aj, 67, 471 
\bibitem[King \& Kiser(1973)]{1973ApJ...181...27K} King, I.~R., \& Kiser,
   J.\ 1973, \apj, 181, 27 
\bibitem[Klimentowski et al.(2009)]{2009MNRAS.397.2015K} Klimentowski, J.,
  {\L}okas, E.~L., Kazantzidis, S., Mayer, L., \& Mamon, G.~A.\ 2009, \mnras,
  397, 2015
\bibitem[Koleva et al.(2009)]{2009MNRAS.396.2133K} Koleva, M., de Rijcke, S.,
  Prugniel, P., Zeilinger, W.~W., \& Michielsen, D.\ 2009, \mnras, 396, 2133 
\bibitem[Koleva et al.(2011)]{2011MNRAS.417.1643K} Koleva, M., Prugniel, P.,
  de Rijcke, S., \& Zeilinger, W.~W.\ 2011, \mnras, 417, 1643 
\bibitem[Koleva et al.(2013)]{2013MNRAS.428.2949K} Koleva, M., Bouchard, A.,
  Prugniel, P., De Rijcke, S., \& Vauglin, I.\ 2013, \mnras, 428, 2949 
\bibitem[Kormendy \& Bender(2012)]{2012ApJS..198....2K} Kormendy, J., \&
  Bender, R.\ 2012, \apjs, 198, 2 
\bibitem[Krajnovi{\'c} et al.(2013)]{2013MNRAS.432.1768K} Krajnovi{\'c}, D.,
  Alatalo, K., Blitz, L., et al.\ 2013, \mnras, 432, 1768 
\bibitem[Lacerna et al.(2016)]{2016A&A...588A..79L} Lacerna, I.,
  Hern{\'a}ndez-Toledo, H.~M., Avila-Reese, V., Abonza-Sane, J., \& del Olmo,
  A.\ 2016, \aap, 588, A79 
\bibitem[Lagos et al.(2015)]{2015MNRAS.448.1271L} Lagos, C.~d.~P., Padilla,
  N.~D., Davis, T.~A., et al.\ 2015, \mnras, 448, 1271 
\bibitem[Lane et al.(2015)]{2015A&A...574A..93L} Lane, R.~R., Salinas, R., \&
   Richtler, T.\ 2015, \aap, 574, A93 
\bibitem[Larson et al.(1980)]{1980ApJ...237..692L} Larson, R.~B., Tinsley,
  B.~M., \& Caldwell, C.~N.\ 1980, \apj, 237, 692 
\bibitem[Laurikainen et al.(2010)]{2010MNRAS.405.1089L} Laurikainen, E.,
   Salo, H., Buta, R., Knapen, J.~H., \& Comer{\'o}n, S.\ 2010, \mnras, 405, 1089
\bibitem[Laurikainen et al.(2011)]{2011MNRAS.418.1452L} Laurikainen, E.,
   Salo, H., Buta, R., \& Knapen, J.~H.\ 2011, \mnras, 418, 1452
\bibitem[Liller(1966)]{1966ApJ...146...28L} Liller, M.~H.\ 1966, \apj, 146, 28 
\bibitem[Lisker et al.(2009)]{2009AN....330..966L} Lisker, T.,
  Brunngr{\"a}ber, R., \& Grebel, E.~K.\ 2009, Astronomische Nachrichten, 330,
  966
\bibitem[Lisker \& Fuchs(2009)]{2009A&A...501..429L} Lisker, T., \& Fuchs,
  B.\ 2009, \aap, 501, 429 
\bibitem[Lisker et al.(2006)]{2006AJ....132..497L} Lisker, T., Grebel, E.~K.,
   \& Binggeli, B.\ 2006, \aj, 132, 497 
\bibitem[{\L}okas et al.(2010)]{2010ApJ...725.1516L} {\L}okas, E.~L.,
  Kazantzidis, S., Majewski, S.~R., et al.\ 2010, \apj, 725, 1516 
\bibitem[Lu et al.(2011)]{2011MNRAS.416..660L} Lu, Y., Kere{\v s}, D., Katz,
  N., et al.\ 2011, \mnras, 416, 660 
\bibitem[Makarova et al.(2016)]{Makarova:2016}Makarova, L.N., Makarov, D.I.,
  Karachentsev, I.D., Tully, R.B., Rizzi, L.\ 2016, MNRAS, 464, 2281
\bibitem[Maller et al.(2006)]{2006ApJ...647..763M} Maller, A.~H., Katz, N.,
  Kere{\v s}, D., Dav{\'e}, R., \& Weinberg, D.~H.\ 2006, \apj, 647, 763 
\bibitem[Mastropietro et al.(2005)]{2005MNRAS.364..607M} Mastropietro, C.,
  Moore, B., Mayer, L., et al.\ 2005, \mnras, 364, 607 
\bibitem[Matkovi{\'c} \& Guzm{\'a}n(2005)]{2005MNRAS.362..289M} Matkovi{\'c},
   A., \& Guzm{\'a}n, R.\ 2005, \mnras, 362, 289 
\bibitem[Mayer et al.(2001a)]{2001ApJ...559..754M} Mayer, L., Governato, F.,
  Colpi, M., et al.\ 2001a, \apj, 559, 754 
\bibitem[Mayer et al.(2001b)]{2001ApJ...547L.123M} Mayer, L., Governato, F.,
  Colpi, M., et al.\ 2001b, \apjl, 547, L123 
\bibitem[M\'endez-Abreu et al.(2016)]{2016arXiv161005324M} M\'endez-Abreu, J.,
   Ruiz-Lara, T., S\'anchez-Menguiano, L., et al.\ 2016, A\&A, 598, A32
\bibitem[Merluzzi et al.(2013)]{2013MNRAS.429.1747M} Merluzzi, P., Busarello,
  G., Dopita, M.~A., et al.\ 2013, \mnras, 429, 1747 
\bibitem[Merluzzi et al.(2016)]{2016MNRAS.460.3345M} Merluzzi, P., Busarello,
  G., Dopita, M.~A., et al.\ 2016, \mnras, 460, 3345 
\bibitem[Merritt(2006)]{2006ApJ...648..976M} Merritt, D.\ 2006, \apj, 648,
   976 
\bibitem[Merritt \& Milosavljevi{\'c}(2005)]{2005LRR.....8....8M} Merritt,
   D., \& Milosavljevi{\'c}, M.\ 2005, Living Reviews in Relativity, 8, 8
\bibitem[Michard \& Marchal(1993)]{1993A&AS...98...29M} Michard, R., \&
  Marchal, J.\ 1993, A\&AS, 98, 29
\bibitem[Michielsen et al.(2008)]{2008MNRAS.385.1374M} Michielsen, D.,
  Boselli, A., Conselice, C.~J., et al.\ 2008, \mnras, 385, 1374 
\bibitem[Monaco et al.(2009)]{2009A&A...502L...9M} Monaco, L., Saviane, I.,
   Perina, S., et al.\ 2009, \aap, 502, L9  
\bibitem[Moore et al.(1996)]{1996Natur.379..613M} Moore, B., Katz, N., Lake,
  G., Dressler, A., \& Oemler, A.\ 1996, \nat, 379, 613
\bibitem[Moore et al.(1998)]{1998ApJ...495..139M} Moore, B., Lake, G., \&
  Katz, N.\ 1998, \apj, 495, 139 
\bibitem[Moorman et al.(2016)]{2016:Moorman}Moorman, C.M., Moreno, J., White,
   A., et al.\ 2016, ApJ, 831, 118 
\bibitem[Mould et al.(2000)]{2000ApJ...529..786M} Mould, J.~R., Huchra, J.~P.,
  Freedman, W.~L., et al.\ 2000, \apj, 529, 786 
\bibitem[Murali et al.(2002)]{2002ApJ...571....1M} Murali, C., Katz, N.,
  Hernquist, L., Weinberg, D.~H., \& Dav{\'e}, R.\ 2002, \apj, 571, 1 
\bibitem[Nieto et al.(1988)]{1988A&A...195L...1N} Nieto, J.-L., Capaccioli,
  M., \& Held, E.~V.\ 1988, A\&A, 195, L1
\bibitem[Norris et al.(2014)]{2014MNRAS.443.1151N} Norris, M.~A., Kannappan,
   S.~J., Forbes, D.~A., et al.\ 2014, \mnras, 443, 1151 
\bibitem[Oh et al.(2016)]{Oh:2016}Oh, S., Yi, S.K., Cortese, L., et al.\ 2016,
  ApJ, 832, 69 
\bibitem[Owers et al.(2012)]{2012ApJ...750L..23O} Owers, M.~S., Couch, W.~J.,
  Nulsen, P.~E.~J., \& Randall, S.~W.\ 2012, \apjl, 750, L23 
\bibitem[Papovich et al.(2016)]{Papovich:2016}Papovich, C., Labb\'e, I.,
   Glazebrook, K., et al.\ 2016, Nature Astronomy, 1, 0003
\bibitem[Patterson(1940)]{Pat40}Patterson, F.S., 1940, Harvard College Observatory Bulletin, 914, 9
\bibitem[Paturel et al.(2003)]{2003A&A...412...45P} Paturel, G., Petit, C.,
  Prugniel, P., et al.\ 2003, \aap, 412, 45 
\bibitem[Paudel et al.(2014)]{2014MNRAS.443..446P} Paudel, S., Lisker, T.,
  Hansson, K.~S.~A., \& Huxor, A.~P.\ 2014, \mnras, 443, 446 
\bibitem[Paudel et al.(2016a)]{2016ApJ...820L..19P} Paudel, S., Hilker, M.,
   Ree, C.~H., \& Kim, M.\ 2016a, \apjl, 820, L19 
\bibitem[Paudel et al.(2016b)]{2016arXiv161103563P} Paudel, S., Lisker, T.,
   Huxor, A.P., Ree, C.H.\ 2016b, MNRAS, 465, 1950 
\bibitem[Paudel et al.(2010)]{2010:Paudel} Paudel, S., Lisker, T.,
  Kuntschner, H., Grebel, E.~K., \& Glatt, K.\ 2010, \mnras, 405, 800 
\bibitem[Paudel \& Ree(2014)]{2014ApJ...796L..14P} Paudel, S., \& Ree,
  C.~H.\ 2014, \apjl, 796, L14 
\bibitem[Pedraz et al.(2002)]{2002MNRAS.332L..59P} Pedraz, S., Gorgas, J.,
  Cardiel, N., S{\'a}nchez-Bl{\'a}zquez, P., \& Guzm{\'a}n, R.\ 2002, \mnras,
  332, L59
\bibitem[Penny \& Conselice(2008)]{PaC08}Penny, S.J., Conselice, C.J.\ 2008,
  MNRAS, 383, 247
\bibitem[Penny et al.(2014)]{2014MNRAS.443.3381P} Penny, S.~J., Forbes, D.~A.,
  Pimbblet, K.~A., \& Floyd, D.~J.~E.\ 2014, \mnras, 443, 3381 
\bibitem[Penny et al.(2016)]{Penny:2016} Penny, S.J., Masters, K.L.,
  Weijmans, A.-M., et al.\ 2016, MNRAS, 462, 3955 
\bibitem[Pichon et al.(2011)]{2011MNRAS.418.2493P} Pichon, C., Pogosyan, D.,
  Kimm, T., et al.\ 2011, MNRAS, 418, 2493
\bibitem[Planck Collaboration et al.(2015)]{2015arXiv150201589P} Planck
  Collaboration, Ade, P.~A.~R., Aghanim, N., et al.\ 2015, A\&A, 594, A13
\bibitem[Pohlen et al.(2004)]{2004ASSL..319..713P} Pohlen, M., Beckman,
   J.~E., H{\"u}ttemeister, S., et al.\ 2004, in Penetrating Bars Through Masks
   of Cosmic Dust, Edited by D.L.\ Block, I.\ Puerari, K.C.\ Freeman,
   R.\ Groess, and E.K.\ Block, Dordrecht: Kluwer Academic Publishers,
   Astrophysics and space science library (ASSL), 319, 713 
\bibitem[Pota et al.(2013)]{2013MNRAS.428..389P} Pota, V., Forbes, D.~A.,
  Romanowsky, A.~J., et al.\ 2013, \mnras, 428, 389 
\bibitem[Prieto et al.(2013)]{2013MNRAS.436.2301P} Prieto, J., Jimenez, R., \&
  Haiman, Z.\ 2013, MNRAS, 436, 2301
\bibitem[Reda et al.(2004)]{2004MNRAS.354..851R} Reda, F.~M., Forbes, D.~A.,
   Beasley, M.~A., O'Sullivan, E.~J., \& Goudfrooij, P.\ 2004, \mnras, 354, 851
\bibitem[Richtler et al.(2015)]{2015A&A...574A..21R} Richtler, T., Salinas,
   R., Lane, R.~R., Hilker, M., \& Schirmer, M.\ 2015, \aap, 574, A21
\bibitem[Roediger et al.(2016)]{Roediger:2016}Roediger, J.C., Ferrarese, L.,
  C\^ot\'e, P., et al.\ 2016, ApJ, 836, 120
\bibitem[Romanowsky et al.(2003)]{2003Sci...301.1696R} Romanowsky, A.~J.,
   Douglas, N.~G., Arnaboldi, M., et al.\ 2003, Science, 301, 1696
\bibitem[Romanowsky \& Fall(2012)]{2012ApJS..203...17R} Romanowsky, A.~J., \&
  Fall, S.~M.\ 2012, \apjs, 203, 17 
\bibitem[Rood(1965)]{1965AJ.....70T.689R} Rood, H.~J.\ 1965, \aj, 70, 689 
\bibitem[Ry\'s et al.(2015)]{2015MNRAS.452.1888R} Ry\'s, A., Koleva, M.,
  Falc{\'o}n-Barroso, J., et al.\ 2015, \mnras, 452, 1888 
\bibitem[S{\'a}nchez et al.(2012)]{2012A&A...538A...8S} S{\'a}nchez, S.~F.,
   Kennicutt, R.~C., Gil de Paz, A., et al.\ 2012, \aap, 538, A8 
\bibitem[Sancisi et al.(2008)]{2008A&ARv..15..189S} Sancisi, R., Fraternali,
  F., Oosterloo, T., \& van der Hulst, T.\ 2008, A\&ARv, 15, 189 
\bibitem[Sandage \& Binggeli(1984)]{1984AJ.....89..919S} Sandage, A., \&
  Binggeli, B.\ 1984, \aj, 89, 919 
\bibitem[Sanduleak \& Pesch(1987)]{1987ApJS...63..809S} Sanduleak, N., \&
  Pesch, P.\ 1987, ApJS, 63, 809 
\bibitem[Saviane et al.(2010)]{2010IAUS..262..426S} Saviane, I., Monaco, L.,
   \& Hallas, T.\ 2010, in Stellar Populations -- Planning for the Next Decade,
   G.Bruzual and S.Charlot (eds), (Cambridge: Cambridge Univ.\ Press), IAU Symp., 262, 426
\bibitem[Savorgnan \& Graham(2016a)]{2016_ApJS_Sav} Savorgnan, G.~A.~D.,
  \& Graham, A.~W.\ 2016a, ApJS, 222, 10 
\bibitem[Savorgnan \& Graham(2016b)]{2016MNRAS.457..320S} Savorgnan, G.~A.~D.,
  \& Graham, A.~W.\ 2016b, \mnras, 457, 320 
\bibitem[Schlafly \& Finkbeiner(2011)]{2011ApJ...737..103S} Schlafly, E.~F.,
  \& Finkbeiner, D.~P.\ 2011, \apj, 737, 103 
\bibitem[Schlegel et al.(1998)]{1998ApJ...500..525S} Schlegel, D.~J.,
  Finkbeiner, D.~P., \& Davis, M.\ 1998, \apj, 500, 525 
\bibitem[Scott et al.(2014)]{2014MNRAS.441..274S} Scott, N., Davies, R.~L.,
   Houghton, R.~C.~W., et al.\ 2014, \mnras, 441, 274 
\bibitem[Seibert et al.(2012)]{2012AAS...21934001S} Seibert, M., Wyder, T.,
  Neill, J., et al.\ 2012, American Astronomical Society Meeting Abstracts, 
  219, 340.01 
\bibitem[Sellwood \& Wilkinson(1993)]{SW93}Sellwood J.A., Wilkinson A.\ 1993,
   Rep.\ Prof.\ Phys., 56, 173
\bibitem[Semelin \& Combes(2005)]{2005A&A...441...55S} Semelin, B., \& Combes,
  F.\ 2005, A\&A, 441, 55 
\bibitem[S{\'e}rsic(1963)]{1963BAAA....6...41S} S{\'e}rsic, J.~L.\ 1963,
  Boletin de la Asociacion Argentina de Astronomia La Plata Argentina, 6, 41 
\bibitem[Sheinis et al.(2000)]{2000SPIE.4008..522S} Sheinis, A.~I., Miller,
  J.~S., Bolte, M., \& Sutin, B.~M.\ 2000, \procspie, 4008, 522 
\bibitem[Simien \& Prugniel(2002)]{2002A&A...384..371S} Simien, F., \&
  Prugniel, P.\ 2002, \aap, 384, 371 
\bibitem[Simien \& Michard(1990)]{Simien:1990}Simien, F., Michard, R.\ 1990,
   A\&A, 227, 11
\bibitem[Smith et al.(2010)]{2010MNRAS.405.1723S} Smith, R., Davies, J.~I.,
   \& Nelson, A.~H.\ 2010, \mnras, 405, 1723 
\bibitem[Smith et al.(2015)]{2015MNRAS.454.2502S} Smith, R.,
  S{\'a}nchez-Janssen, R., Beasley, M.~A., et al.\ 2015, \mnras, 454, 2502 
\bibitem[Stewart et al.(2013)]{2013ApJ...769...74S} Stewart, K.~R., Brooks,
   A.~M., Bullock, J.~S., et al.\ 2013, ApJ, 769, 74
\bibitem[Stott et al.(2007)]{2007ApJ...661...95S} Stott, J.~P., Smail, I.,
  Edge, A.~C., et al.\ 2007, \apj, 661, 95 
\bibitem[Teodorescu et al.(2005)]{2005ApJ...635..290T} Teodorescu, A.~M.,
   M{\'e}ndez, R.~H., Saglia, R.~P., et al.\ 2005, \apj, 635, 290 
\bibitem[Toloba et al.(2014)]{2014ApJ...783..120T} Toloba, E., Guhathakurta,
  P., van de Ven, G., et al.\ 2014, \apj, 783, 120 
\bibitem[Toloba et al.(2015)]{2015ApJ...799..172T} Toloba, E., Guhathakurta,
  P., Boselli, A., et al.\ 2015, \apj, 799, 172 
\bibitem[Toloba et al.(2016)]{2016:Toloba} Toloba, E., Li, B., Guhathakurta,
  P.\ 2016, ApJ, 822, 51 
\bibitem[van de Voort et al.(2011)]{2011MNRAS.414.2458V} van de Voort, F.,
   Schaye, J., Booth, C.~M., Haas, M.~R., \& Dalla Vecchia, C.\ 2011, \mnras,
  414, 2458 
\bibitem[van den Bergh(1976)]{1976ApJ...206..883V} van den Bergh, S.\ 1976,
  \apj, 206, 883 
\bibitem[van der Kruit(1987)]{1987A&A...173...59V} van der Kruit, P.~C.\ 1987,
  \aap, 173, 59 
\bibitem[Weiner et al.(2006)]{2006ApJ...653.1027W} Weiner, B.~J., Willmer,
   C.~N.~A., Faber, S.~M., et al.\ 2006, \apj, 653, 1027 
\bibitem[Weinmann et al.(2011)]{2011MNRAS.416.1197W} Weinmann, S.~M., Lisker,
  T., Guo, Q., Meyer, H.~T., \& Janz, J.\ 2011, \mnras, 416, 1197
\bibitem[White \& Frenk(1991)]{1991ApJ...379...52W} White, S.~D.~M., \&
   Frenk, C.~S.\ 1991, ApJ, 379, 52
\bibitem[Wirth \& Gallagher(1984)]{WaG84}Wirth, A., Gallagher, J.S.\ 1984, ApJ, 282, 85
\bibitem[Woodley et al.(2007)]{2007AJ....134..494W} Woodley, K.~A., Harris,
   W.~E., Beasley, M.~A., et al.\ 2007, \aj, 134, 494 
\bibitem[Wright(2006)]{2006PASP..118.1711W} Wright, E.~L.\ 2006, \pasp, 118,
  1711 
\bibitem[Wright et al.(2010)]{2010AJ....140.1868W} Wright, E.~L., Eisenhardt,
   P.~R.~M., Mainzer, A.~K., et al.\ 2010, \aj, 140, 1868
\bibitem[Yagi et al.(2010)]{2010AJ....140.1814Y} Yagi, M., Yoshida, M.,
  Komiyama, Y., et al.\ 2010, \aj, 140, 1814 
\bibitem[Zaritsky et al.(2008)]{2008ApJ...682...68Z} Zaritsky, D., Zabludoff,
   A.~I., \& Gonzalez, A.~H.\ 2008, \apj, 682, 68
\end{thebibliography}
\end{document}